\documentclass[aps,pra,10pt,showpacs,twocolumn,superscriptaddress,longbibliography]{revtex4-1}
\usepackage{graphicx}
\usepackage[usenames]{color}
\usepackage{amssymb,amsmath}
\usepackage{physics}
\usepackage{siunitx}
\usepackage{xcolor}
\usepackage{setspace} 
\usepackage{float} 
\usepackage{tabularx}
\usepackage[linkcolor=blue,citecolor=blue,colorlinks=true,urlcolor=blue]{hyperref}

\newcommand{\fig}[1]{Fig.~\ref{#1}}
\newcommand{\be}[1]{\begin{equation}\label{#1}}
\newcommand{\ee}{\end{equation}}
\usepackage{amsmath,esint}
\usepackage{comment}
\usepackage{lipsum}
\usepackage[toc]{appendix}

\begin{document}

\title{Nondipole electron momentum offset as a probe of correlated three electron ionization in strongly driven atoms}

\author{G. P. Katsoulis}
\affiliation{Department of Physics and Astronomy, University College London, Gower Street, London WC1E 6BT, United Kingdom}
\author{M. B. Peters}
\affiliation{Department of Physics and Astronomy, University College London, Gower Street, London WC1E 6BT, United Kingdom}
\author{A. Emmanouilidou}
\affiliation{Department of Physics and Astronomy, University College London, Gower Street, London WC1E 6BT, United Kingdom}

\begin{abstract}
We employ a recently developed three-dimensional semiclassical model to identify nondipole effects in triple  ionization of Ne driven by infrared laser pulses at intensities where electron-electron correlation prevails. This model fully  accounts for the Coulomb interaction of each electron with the core and avoids artificial autoionization by employing effective Coulomb potentials to describe the interaction between bound electrons (ECBB). Using the ECBB model, we identify a prominent signature of nondipole effects. Namely, the component along the direction of light propagation of the average sum of the final electron momenta is large and positive. That is, we identify a positive momentum offset, absent in the dipole approximation. We find that this positive momentum offset  stems mostly from the momentum change due to the magnetic field. To further understand this  momentum change,  we also develop a simple model for the motion of an electron inside an electromagnetic field. This simple model accounts for the effect of the Coulomb forces only as a sharp change in the momentum of the electron during recollision.  We show that the momentum change due to the magnetic field is related with the sharp change in momentum during recollision for the recolliding electron as well as with the time of recollision for both the recolliding and bound electrons. Hence, we demonstrate  that the final electron momentum offset  probes the strength of a recollision and hence  the degree of  correlation in multielectron ionization.
 \end{abstract}
\date{\today}

\maketitle

\section{Introduction}
 Nonsequential multielectron ionization (NSMI) in atoms driven by intense and infrared  laser pulses is a fundamental process  underlied by electron-electron correlation \cite{knee1983}. The theoretical study of multielectron ionization of strongly driven systems constitutes a big computational challenge.  Taking also into account the spatial dependence of the vector potential $\mathbf{A}(\mathbf{r},t)$ and consequently accounting for the magnetic field, $\mathbf{B}(\mathbf{r},t) = \nabla \times\mathbf{A}(\mathbf{r},t),$ adds to the computational difficulty. Hence, most theoretical studies are formulated in the dipole approximation. However, to fully explore ionization phenomena and identify nondipole effects in driven atoms and molecules one needs to account for the Lorentz force $\mathbf{F_B}=q\mathbf{v}\times \mathbf{B}$ exerted on particles of charge $q$ moving with velocity $\mathbf{v}$. 
 
 Magnetic fields effects have been previously identified in a wide range of processes. For example, in stabilization \cite{stabilization}, in high-harmonic generation \cite{HHG1,HHG2,HHG3}, and in multielectron ionization probabilities of Ne \cite{Neon}, with observable effects  found only for intensities two orders of magnitude  larger than the ones considered in the current work. For the largest intensity we consider here, we find that the amplitude $\beta_0\approx U_p / \left( 2 \omega c \right)$ of the electron motion due to $\mathbf{F_B}$ is roughly 0.2 a.u., instead of the expected 1 a.u.  \cite{Magnetic1,Magnetic2}, where  $U_p$ is the ponderomotive energy.
 Over the last years, there has been an intense interest in nondipole effects \cite{PhysRevLett.106.193002,KellerMagnetic2014,CorkumBandrauk2015,Biegert2015,Emmanouilidou1,Emmanouilidou2,KellerMagneticFields,FSun,BurgdorferPRL2022,PhysRevLett.128.113201}. Advanced studies \cite{Emmanouilidou1,Emmanouilidou2} have predicted  nondipole effects in correlated two-electron ionization, which have been  verified experimentally for driven Ar  \cite{FSun}. Nondipole effects in nonsequential double ionization were also studied in a recent experiment on strongly driven Xe \cite{PhysRevLett.128.113201}. 

We have previously reported \textit{nondipole gated double ionization}, a prominent mechanism of nondipole effects in nonsequential double ionization of strongly driven atoms \cite{Emmanouilidou1,Emmanouilidou2}. The magnetic field jointly with a  recollision act as a gate that allows for double ionization to occur only for a subset of the initial momenta of the recolliding electron along the direction of light propagation. Namely, the recolliding electron has an average initial momentum that is negative along the direction of light propagation ($y$ axis), while it is zero in the dipole approximation. This negative initial momentum compensates for the positive momentum shift induced by the Lorentz force, allowing for the recolliding electron to return to the core. As a result, the recolliding electron just before recollision arrives from the $-y$ axis with positive momentum. For the case of strongly driven He at high intensities, we have shown that the recollisions involved are glancing ones. As a result, the recolliding electron just before recollision is accelerated by the Coulomb attraction from the core resulting in the $y$ component of the average sum of the final electron momenta being large and positive \cite{Emmanouilidou1,Emmanouilidou2}. 

For triple ionization of Ne, for intensities where strong  and not glancing recollisions prevail, we find that  nondipole gated ionization is still present, i.e., the recolliding electron has a  negative avaerage initial momentum. We demonstrate that the strong recollisions involved for driven Ne result in a different physical mechanism, compared to driven He, giving rise to a large positive $y$ component of the average sum of the final electron momenta.
 
 For driven Ne, using the ECBB model, we identify the change in momentum due to the magnetic field as the main source for the positive momentum offset along the $y$ axis. 
 To better understand this momentum change,  we also develop a simple model to describe the motion of an electron inside an electromagnetic field. In this simple model, we take into account the effect of the Coulomb forces via  a sharp change in the momentum of each electron during recollision. Using this model,  we show that for the recolliding electron the value of the positive momentum offset has a one-to-one correspondence with the magnitude of the momentum change along the $y$ axis during recollision. That is, a strong recollision results in a large positive offset. For a bound and a recolliding electron, we also find  that the value of the momentum offset depends on the time of recollision. Namely, a strong recollision that takes place around a zero of the electric field results in a large positive momentum offset. 
 
 Hence, in this work we demonstrate that the positive momentum offset probes the strength of the recollision involved and hence of the degree of the resulting correlated multielectron ionization.
We show this by our finding of a larger positive momentum offset for triple compared to double ionization and for the direct compared to the delayed recollision pathway of driven Ne. Indeed, we show that triple ionization is more correlated than double ionization and that electron-electron dynamics is more correlated in the direct compared to the delayed pathway. 
\section{Method}
In this work, we identify nondipole effects in triple and double ionization of Ne. To do so, we employ a three-dimensional (3D) semiclassical model that employs effective Coulomb potentials to describe the interaction between bound electrons (ECBB) \cite{Agapi3electron,AgapiNeonPRL}. This model is developed in the nondipole framework. Moreover, this model was developed to address the main challenge that classical and semiclassical models of NSMI face, i.e., unphysical autoionization. Due to the singularity in the Coulomb potential, one of the bound electrons can undergo a close encounter with the core and acquire very negative energy. This leads to the escape of another bound electron via the Coulomb interaction between bound electrons. One way to avoid this is by softening the Coulomb potential, see  Refs. \cite{PhysRevLett.97.083001,Zhou:10,Tang:13} for nonsequential triple ionization. Alternatively, Heisenberg potentials (effective softening)  \cite{PhysRevA.21.834} are added that  mimic the Heisenberg uncertainty principle and prevent  each electron from a close encounter with the core \cite{PhysRevA.104.023113,PhysRevA.105.053119}. 

However, softening the Coulomb potential fails to accurately describe electron scattering from the core \cite{Pandit2017,Pandit2018}, rendering the softened potentials quite inaccurate for high energy recolliding electrons. In contrast, in the ECBB model, we treat exactly the Coulomb singularity in the interaction of an electron with the core as well as the interaction between a quasifree and a bound electron. Here, quasifree refers to a recolliding electron or an electron escaping to the continuum.  To address the autoionization problem, since we treat the electron-core interaction accurately, we use an effective Coulomb potential to describe the interaction between two bound electrons. We have shown that treating accurately the electron-core interaction is of paramount importance in obtaining accurate ionization spectra \cite{AgapiNeonPRL}. Indeed, using the ECBB model, we have investigated three-electron ionization in Ar \cite{Agapi3electron} and Ne \cite{AgapiNeonPRL} driven by infrared pulses. For triple ionization, we have shown that the probability distribution of the sum of the final electron momenta along the $z$ axis is in very good agreement with experimental results, especially for Ne \cite{AgapiNeonPRL}. In this work, the direction of the electric field is along the $z$ axis.

In the ECBB model, we determine on the fly whether an electron is quasifree or bound using the following simple criteria. A quasifree electron can  transition to bound following a recollision. Specifically, once the quasifree electron has its closest encounter with the core, this electron transitions to bound if its position  along the $z$ axis is  influenced more by the core than the electric field. Also, a bound electron  transitions to quasifree due to transfer of energy  during a recollision  or from the laser field.  In the former case, this transition occurs if the potential energy of  this bound electron with the core is constantly decreasing. In the latter case,  if  the energy of the bound electron  becomes  positive and remains positive it transitions to quasifree. The criteria are discussed in detail and  illustrated in \cite{Agapi3electron,AgapiNeonPRL}.

 In our model, one electron tunnel ionizes through the field-lowered Coulomb barrier at time $t_0$. The tunneling occurs with a rate described by the instantaneous quantum-mechanical Ammosov-Delone-Krainov (ADK) formula  \cite{Landau,Delone:91}, adjusted accordingly to account for the depletion of the initial ground state \cite{AgapiNeonPRL}. We find $t_0$ in the time interval [-2$\tau$, 2$\tau$] where the electric field is nonzero, using importance sampling \cite{ROTA1986123}; $\tau$ is the full width at half maximum of the pulse duration in intensity. The exit point of the recolliding electron along the direction of the electric field is obtained  analytically using parabolic coordinates \cite{HUP1997533}. The electron momentum along the electric field is set equal to zero, while the transverse one is given by a Gaussian distribution. This distribution represents the Gaussian-shaped filter with an intensity-dependent width arising from standard tunneling theory \cite{Delone:91,Delone_1998,PhysRevLett.112.213001}. For the initially bound electrons, we employ  a microcanonical distribution  \cite{Agapi3electron}, while the core is initiated at rest at the origin.

We use a vector potential of the form
\begin{equation}\label{eq:vector_potential}
\mathbf{A}(y,t) = -\frac{E_0}{\omega}\exp \left[ - 2\ln (2)\left( \frac{c t - y}{c \tau} \right)^2 \right]   \sin ( \omega t  - k y) \hat{\mathbf{z}},
\end{equation}
where $k=\omega/c$ is the  wave number of the laser field.  The direction of the vector potential and the electric field, $\mathbf{E}(y,t) = -\frac{\partial\mathbf{A}(y,t)}{\partial t}$, is along the $z$ axis, while the direction of light propagation is along the $y$ axis.  The magnetic field, $ \mathbf{B}(y,t) = \grad \times \mathbf{A}(y,t)$, points  along the $x$ axis.  The pulse duration is $\tau = 25$ fs, while  the wavelength is  800 nm. We consider intensities of 1.0, 1.3 and $\mathrm{1.6 \; PW/cm^2 }$. The  highest intensity  considered is chosen such that  the probability for a second electron to tunnel ionize in Ne solely due to the laser field is very small  \cite{AgapiNeonPRL}. Hence, here, electron-electron  correlation prevails in triple and double ionization. In what follows, triple ionization (TI) refers to nonsequential triple ionization (NSTI) and double ionization (DI) to nonsequential double ionization (NSDI). 

 The Hamiltonian of the four-body system is given by
\begin{equation}\label{Hamiltonian_effective}
\begin{aligned}
&H = \sum_{i=1}^{4}\frac{\left[\mathbf{\tilde{p}}_{i}- Q_i\mathbf{A}(y,t) \right]^2}{2m_i}+\sum_{i=2}^{4}\frac{Q_iQ_1}{|\mathbf{r}_1-\mathbf{r}_i|} \\
&+\sum_{i=2}^{3}\sum_{j=i+1}^{4} \left[ 1-c_{i,j}(t)\right]\frac{Q_iQ_j}{|\mathbf{r}_i-\mathbf{r}_j|} +\sum_{i=2}^{3}\sum_{j=i+1}^{4}c_{i,j}(t) \\\Big[ 
& V_{\mathrm{eff}}(\zeta_j(t),|\mathbf{r}_{1}-\mathbf{r}_{i}|) + V_{\mathrm{eff}}(\zeta_i(t),|\mathbf{r}_{1}-\mathbf{r}_{j}|)\Big],
\end{aligned}
\end{equation}
where $Q_i$ is the charge, $m_i$ is the mass, $\mathbf{r}_{i}$ is the position vector and $\mathbf{\tilde{p}}_{i}$ is the canonical momentum vector of particle $i$. The mechanical momentum $\mathbf{p}_{i}$ is given by
\begin{equation}
\mathbf{p}_{i}=\mathbf{\tilde{p}}_{i}-Q_i\mathbf{A}(y,t).
\end{equation}
In the ECBB model, all electrons and the core are allowed to move. The effective Coulomb potential that an electron $i$ experiences at a distance $|\mathbf{r}_{1}-\mathbf{r}_{i}|$ from the core (particle 1),  due to the charge distribution of electron $j$ is derived as follows \cite{PhysRevA.40.6223,Agapi3electron}. We approximate the wavefunction of a bound electron $j$ with a 1s hydrogenic wavefunction
\begin{equation}
\psi(\zeta_j,|\mathbf{r}_{1}-\mathbf{r}_{j}|) = \left( \frac{\zeta_j^3}{\pi} \right)^{1/2} e^{-\zeta_j |\mathbf{r}_{1}-\mathbf{r}_{j}|},
\end{equation} 
with $\zeta_j$ the effective charge of particle $j$ \cite{Agapi3electron,PhysRevA.40.6223}. Hence, using Gauss's law \cite{PhysRevA.40.6223,Agapi3electron}, one finds that the potential produced due to the charge distribution $-|\psi(\zeta_j,|\mathbf{r}_{1}-\mathbf{r}_{j}|) |^2$ is given by
\begin{equation}
V_{\mathrm{eff}}(\zeta_j,|\mathbf{r}_{1}-\mathbf{r}_{i}|) =  \dfrac{1-(1+\zeta_j| \mathbf{r}_{1}-\mathbf{r}_{i}|)e^{-2\zeta_j| \mathbf{r}_{1}-\mathbf{r}_{i}|}}{| \mathbf{r}_{1}-\mathbf{r}_{i}|}.
\end{equation}
When electron $i$ approaches the core, i.e., $|\mathbf{r}_{1}-\mathbf{r}_{i}|\to 0$, the effective potential is equal to $\zeta_j.$ This ensures that the energy transfer between bound electrons is finite and therefore autoionization is prevented. The functions $c_{i,j}(t)$ determine whether at time $t$, during propagation, the full Coulomb potential or the effective $V_{\mathrm{eff}}(\zeta_i,|\mathbf{r}_{1}-\mathbf{r}_{j}|)$ and $V_{\mathrm{eff}}(\zeta_j,|\mathbf{r}_{1}-\mathbf{r}_{i}|)$ potentials describe the interaction between a  pair of electrons $i$ and $j$  \cite{Agapi3electron}. The effective potentials are activated on the fly only when both electrons are bound. 

During time propagation, to accurately account for the Coulomb singularity, we transform the position and momenta  using the global regularization scheme \cite{Heggie1974,Agapi3electron}, first introduced for the gravitational N-body problem \cite{Heggie1974}. We propagate in time the transformed position and momenta of all particles using  the classical ECBB Hamiltonian, see \cite{Agapi3electron}. To propagate, we use a leapfrog technique that allows to solve Hamilton's equations when the derivative of the position and the momentum depends on the quantities themselves \cite{Pihajoki2015,Liu2016,PhysRevA.103.033115}. This technique is employed jointly with the Bulirsch-Stoer method \cite{press2007numerical,bulirsch1966numerical}. 

We stop the  propagation at time $t_f,$ when the energy of each particle converges. We label the trajectory as  triply or  doubly ionized  if three or two electrons have  positive energy,  and compute the triple ionization and double ionization probabilities out of all events. After we label an event as triple or double ionization, we identify the main pathways of energy transfer via the recollision, i.e., we characterise an event as direct or delayed. We explain in detail how to identify the pathways of  triple and double ionization in Refs. \cite{Agapi3electron,AgapiNeonPRL}. In this work we employ atomic units, unless otherwise stated.
   
\section{Results}
\subsection{Correlated electron momenta in TI and DI}\label{Section:Correlated_momenta}
\begin{figure}[b]
\centering
\includegraphics[width=\linewidth]{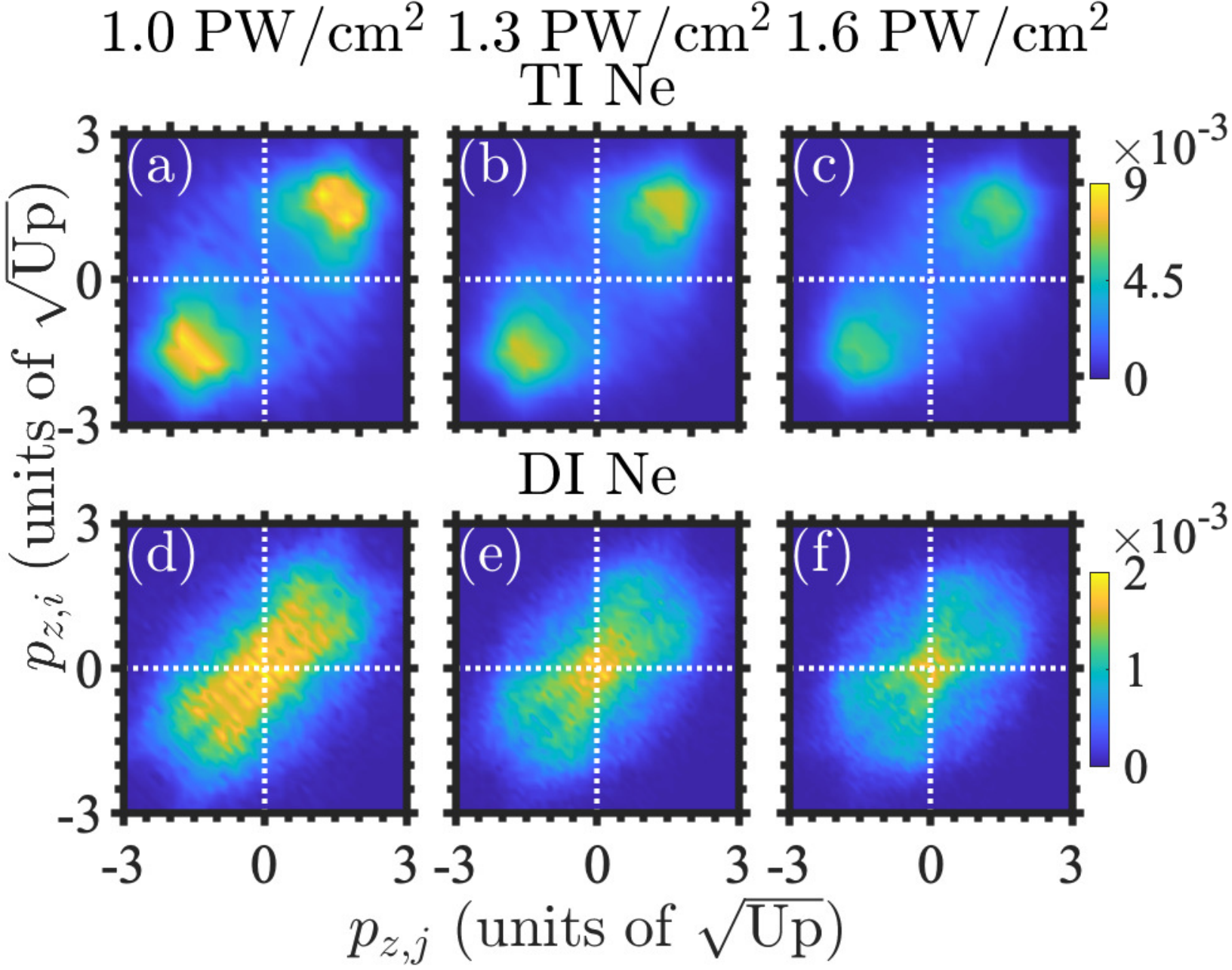}
\caption{For Ne, symmetrized correlated momenta $p_z$ of all three pairs of escaping electrons for triple ionization (top row) and the one pair of escaping electrons for double ionization (bottom row).}\label{fig:Correlated_momenta}
\end{figure}
In \fig{fig:Correlated_momenta}, for TI and DI of driven Ne, we plot the symmetrised correlated electron momenta along the direction of the electric field ($p_z$) for all pairs of escaping electrons. We find that electron-electron dynamics is more correlated in triple compared to double ionization. Specifically, for TI, at all three intensities, we find (not shown) that recollisions occur around a zero of the electric field and hence at an extremum of the vector potential. This results in large final electron momenta in TI, since $p_z$ is roughly equal to minus the vector potential at the  time of recollision. Indeed, this is seen in Figs. \ref{fig:Correlated_momenta}(a)-\ref{fig:Correlated_momenta}(c), where we have a large concentration of electrons with large momenta in the first and third quadrants. Comparing Figs. \ref{fig:Correlated_momenta}(a)-\ref{fig:Correlated_momenta}(c) for TI with Figs. \ref{fig:Correlated_momenta}(d)-\ref{fig:Correlated_momenta}(f) for DI, we find that electron-electron dynamics is more correlated for TI. This is particularly the case at the higher intensity, 1.6 $\mathrm{PW/cm^2}$, where we find that for DI  recollisions occur  around an extremum of the field, i.e., a zero of the vector potential. This results in smaller final electron momenta for DI.

\begin{figure}[t]
\centering
\includegraphics[width=\linewidth]{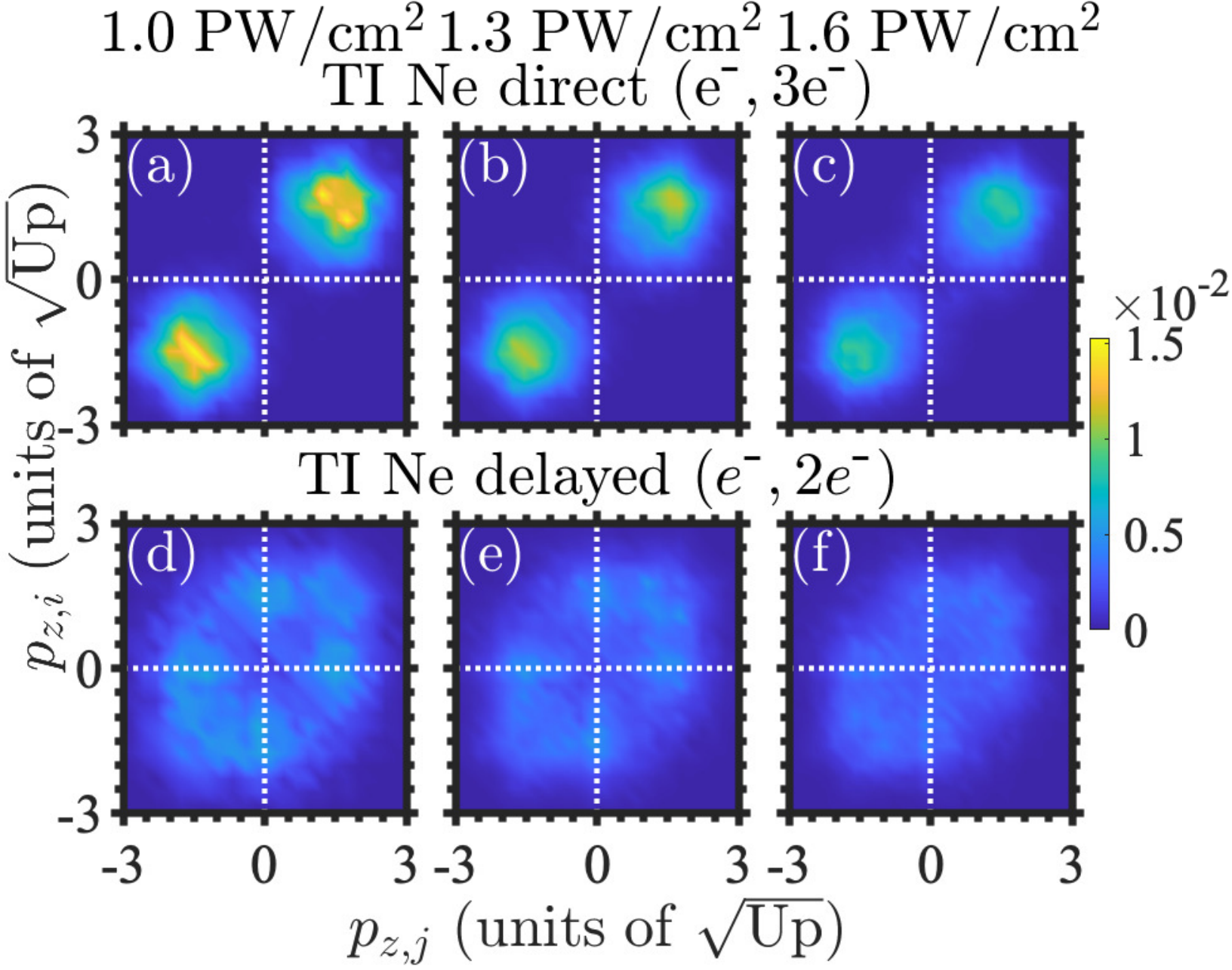}
\caption{For triple ionization of Ne, symmetrized correlated momenta $p_z$ for the direct $(e^{\text{-}},3e^\text{-})$ pathway  (top row) and for the delayed $(e^{\text{-}},2e^\text{-})$ pathway  (bottom row). \label{fig:Correlated_momenta_TI_pathways}}
\end{figure}
\begin{figure}[b]
\centering
\includegraphics[width=\linewidth]{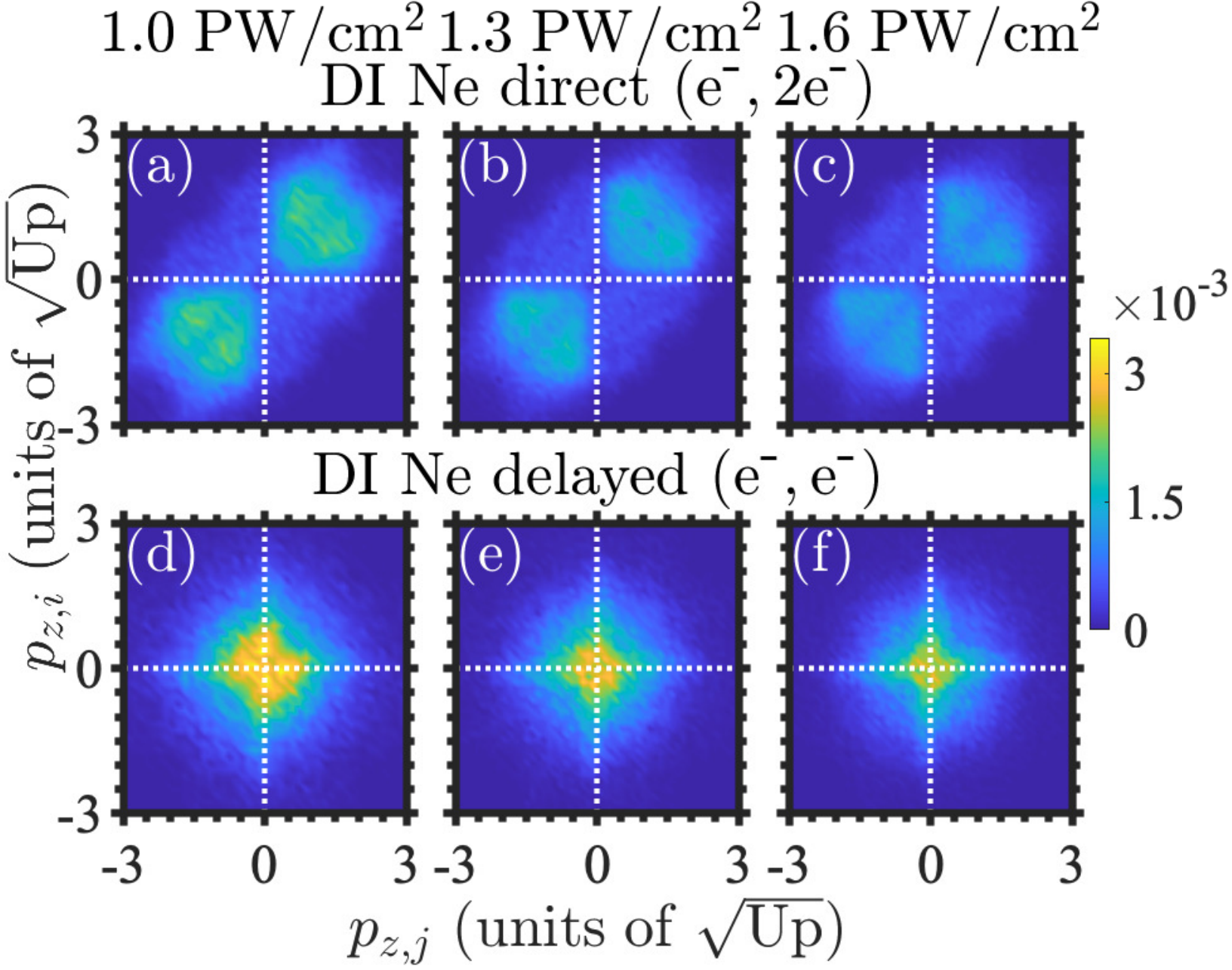}
\caption{For double ionization of Ne, symmetrized correlated momenta for the direct  $(e^{\text{-}},2e^\text{-})$ pathway (top row) and for the delayed $(e^{\text{-}},e^\text{-})$ pathway  (bottom row).}\label{fig:Correlated_momenta_DI_pathways}
\end{figure}

Next, for TI and DI, we show that electron-electron dynamics is more correlated for recollision pathways where more electrons ionize soon after recollision. For TI, we find that the prevailing recollision pathways are the direct $(e^{\text{-}},3e^\text{-})$ and the delayed $(e^{\text{-}},2e^\text{-})$. The notation $(e^{\text{-}},ne^\text{-})$ denotes $n$ electrons ionizing shortly after recollision. Also, DI proceeds mainly via the direct $(e^{\text{-}},2e^\text{-})$ and the delayed $(e^{\text{-}},e^\text{-})$ pathways. We plot the symmetrized correlated electron momenta $p_z$ for the prevailing recollision pathways for TI in \fig{fig:Correlated_momenta_TI_pathways} and for DI in \fig{fig:Correlated_momenta_DI_pathways}. In \fig{fig:Correlated_momenta_TI_pathways}, we clearly show  that electron-electron correlation is higher in the direct pathway [Figs. \ref{fig:Correlated_momenta_TI_pathways}(a)-\ref{fig:Correlated_momenta_TI_pathways}(c)] compared to the delayed pathway [Figs. \ref{fig:Correlated_momenta_TI_pathways}(d)-\ref{fig:Correlated_momenta_TI_pathways}(f)]. Indeed, in \fig{fig:Correlated_momenta_TI_pathways}, for TI, for the direct pathway the majority of events are concentrated in the first and third quadrants while for the delayed pathway the events are more spread out. For DI, \fig{fig:Correlated_momenta_DI_pathways} clearly shows that electron-electron correlation is higher in the direct compared to the delayed pathway at all three intensities. Indeed, for DI, events for the direct pathway are concentrated in the first and third quadrants while for the delayed pathway events are concentrated around zero momentum. Also, we see that electron-electron correlation is higher for the direct pathway of TI compared to the one for DI as well as for the delayed pathway of TI compared to the one of DI.

\subsection{Positive momentum offset in TI}\label{Section:Offset_TI}
\begin{figure}[b]
\centering
\includegraphics[width=\linewidth]{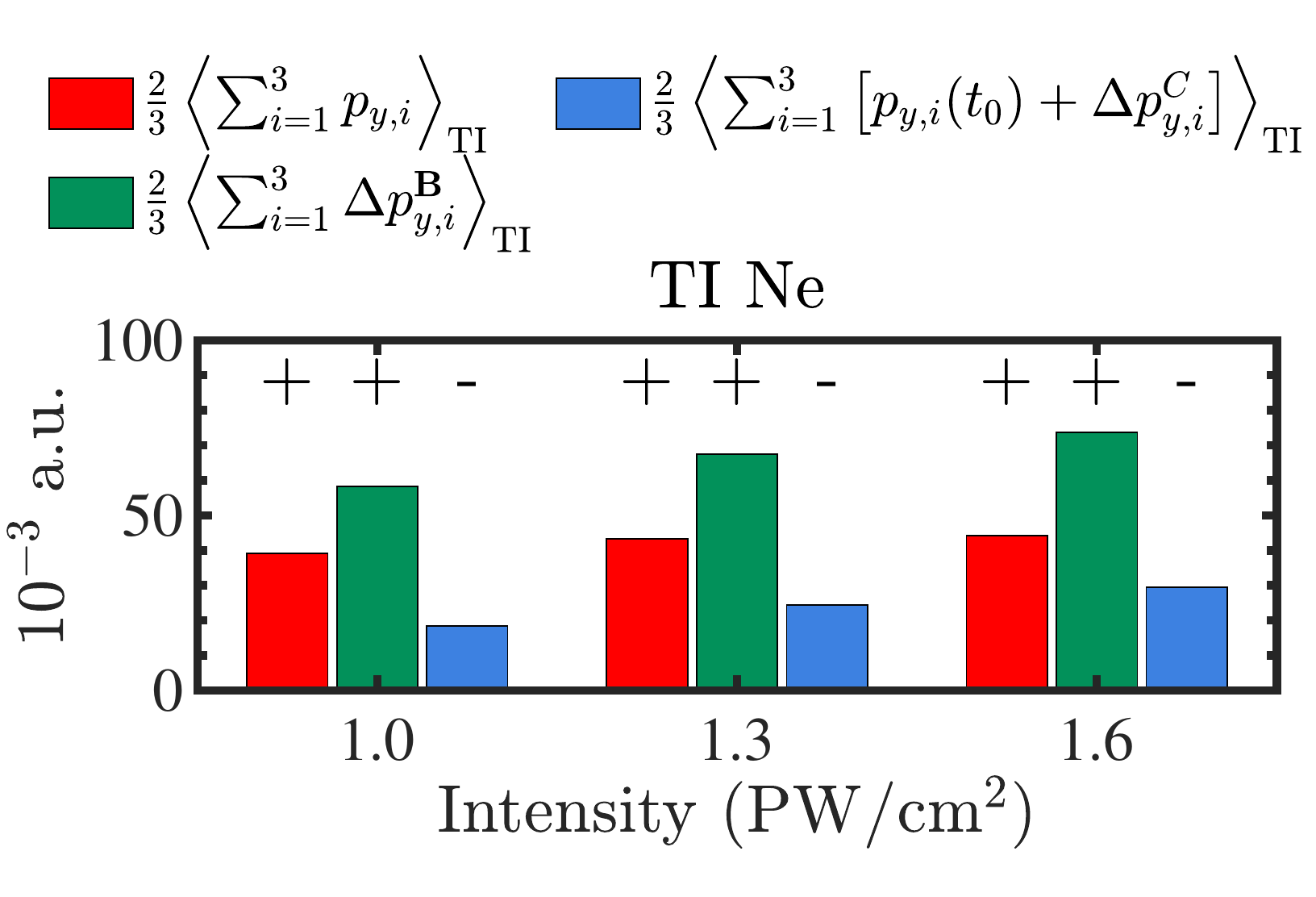}
\caption{For Ne, at each intensity, height of red bar denotes the momentum offset per pair of electrons for TI $2/3\left\langle \sum_{i=1}^{3} p_{y,i} \right\rangle$,
and the contributions due to the magnetic field $2/3\left\langle \sum_{i=1}^{3} \Delta p_{y,i}^{\mathbf{B}} \right\rangle$ (green bar) and due to the initial momentum and Coulomb forces $2/3\left\langle \sum_{i=1}^{3} \left[ p_{y,i}(t_0)+\Delta p_{y,i}^{C} \right]\right\rangle$ (blue bar). The plus (+), minus (-) sign above the bar denotes a positive or negative value, respectively, for the given contribution.}\label{fig:Offset_all_events_TI}
\end{figure}
In \fig{fig:Offset_all_events_TI}, for TI of driven Ne, to obtain the momentum offset per pair of ionizing electrons, we compute the $y$ component (direction of light propagation) of the average sum of the final electron momenta and we then multiply by a factor of 2/3 as follows
\begin{align}\label{eq:offset_TI}
\begin{split}
&\frac{2}{3}\left\langle\sum_{i=1}^{3} p_{y,i}  \right\rangle_{\text{TI}} = \\
&\left\langle \dfrac{ (p_{y,1}+p_{y,2})+(p_{y,1}+p_{y,3}) + (p_{y,2}+p_{y,3})}{3} \right\rangle. 
\end{split}
\end{align}
We denote by $ p_{y,i}$ the $y$ component of the final momentum of electron $i$. The reason we compute for TI the momentum offset per pair of electrons is to directly compare  with the momentum offset in DI where there is only one pair of ionizing electrons. Note that the momentum offset for both TI and DI is zero in the dipole approximation. For TI, the momentum offset is denoted by the height of the red bars in \fig{fig:Offset_all_events_TI}. At all three intensities, we find that the momentum offset has a significant positive value around 0.04 a.u. We find that this is roughly four times larger than twice (to account for an electron pair) the momentum offset in single ionization. 

Next, we identify the reason for the positive value of the momentum offset. To do so, we write the average value of the final electron momentum, $\langle p_{y,i} \rangle$, in terms of three contributions as follows 
\begin{equation}\label{eq:offset_terms}
\langle p_{y,i} \rangle = \langle p_{y,i} (t_0) \rangle  + \left\langle \Delta p^{C}_{y,i} \right\rangle + \left\langle \Delta p^{\mathbf{B}}_{y,i} \right\rangle.
\end{equation}
The first term, $\langle p_{y,i} (t_0) \rangle,$ is the $y$ component of the average value of the initial electron momentum. The next term, $\left\langle \Delta p^{C}_{y,i} \right\rangle$, denotes the $y$ component of the average change in the momentum of electron $i$ in the time interval $[t_0,t_f]$ due to the Coulomb forces and the effective potentials while the term, $\left\langle \Delta p^{\mathbf{B}}_{y,i} \right\rangle$, denotes the corresponding momentum change due to the the magnetic field. \fig{fig:Offset_all_events_TI} clearly shows that the positive momentum offset per pair of electrons for TI is due to the momentum change  from the magnetic field (green bars). \fig{fig:Offset_all_events_TI} also shows that the momentum change due to the Coulomb forces (blue bars) has a negative value and is significantly less compared to the momentum change due to the magnetic field. Hence, in what follows, we only focus on the momentum change due to the magnetic field.

\begin{figure}[t]
\centering
\includegraphics[width=\linewidth]{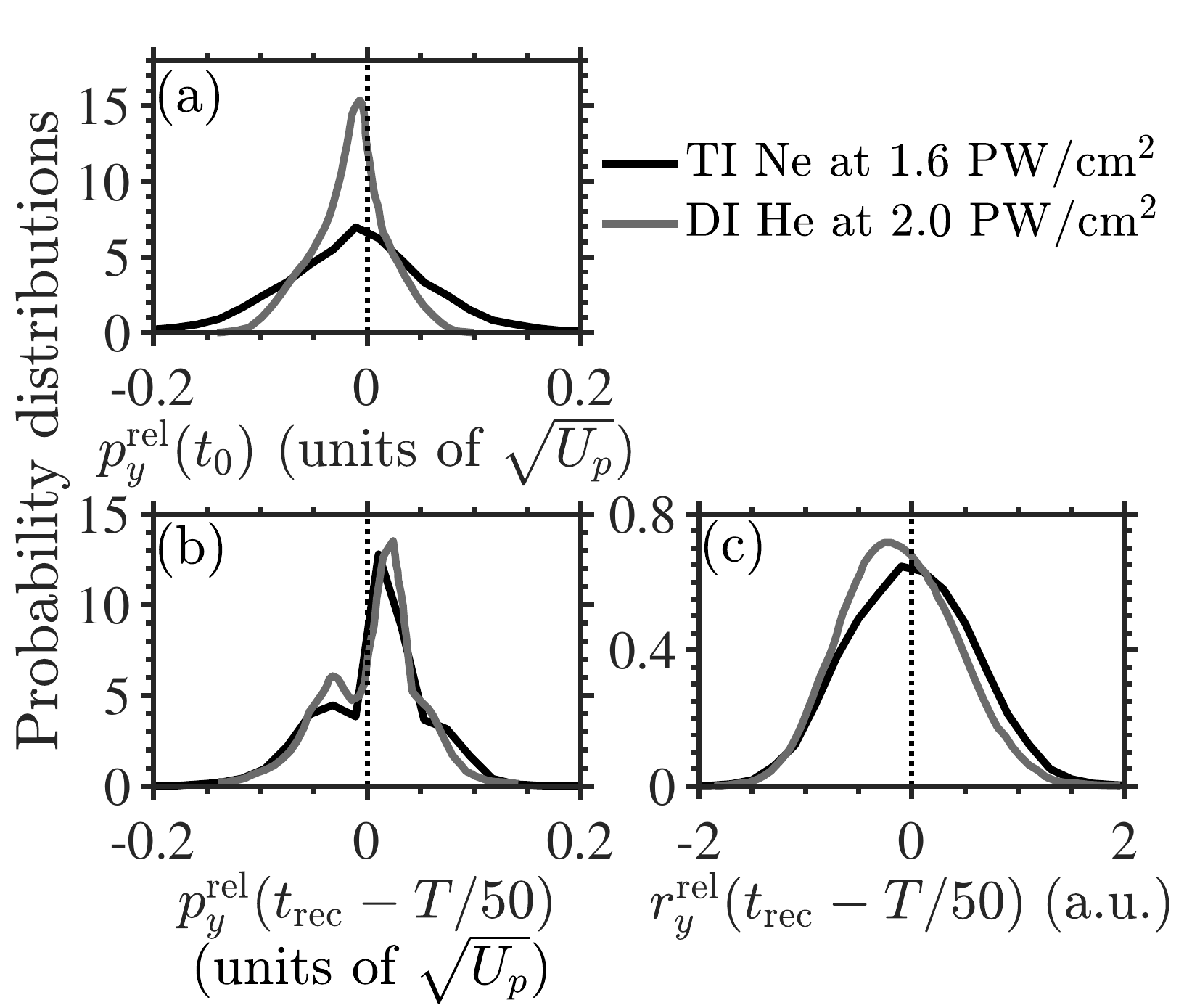}
\caption{Plots of the distribution of the $y$ component of the relative with respect to the core momentum
of the recolliding electron at the time of tunnelling $t_0$ (a), shortly before recollision at time $t_{\mathrm{rec}} - T/50$ (b), and of the $y$ component of the relative with respect to the core position
of the recolliding electron shortly before recollision at time $t_{\mathrm{rec}} - T/50$ (c) for TI of driven Ne at 1.6 $\mathrm{PW/cm^2}$ (black lines) and for DI of driven He at 2 $\mathrm{PW/cm^2}$ (gray lines). $T$ is the period of the laser field. Note that in the study for He the core was kept fixed at the origin\cite{Emmanouilidou2}, while for Ne the core is also moving.} \label{fig:Momenta_and_position_at_trec}
\end{figure}

The above show that for NSTI in driven Ne the mechanism responsible for the positive momentum offset along the $y$ axis is different from nondipole gated ionization identified in strongly driven He \cite{Emmanouilidou1}. In the latter case, the significant positive momentum offset in DI was due to the recolliding electron coming in just before recollision mostly from the $-y$  direction with positive momentum $p_y$ (\fig{fig:Momenta_and_position_at_trec}) and the Coulomb attraction from the core acting to increase $p_y$. However, the recollisions involved in driven He were glancing ones. For NSTI in driven Ne, we find  that the recolliding electron also has a negative average initial momentum along the y axis and also approaches mostly from the $-y$ axis with positive $p_y$ momentum (\fig{fig:Momenta_and_position_at_trec}). However, the recollisions in driven Ne are strong ones resulting in the most important contribution to  the $y$ component of the momentum change being due to the magnetic field and not due to the Coulomb attraction from the core.

\subsection{Momentum change along the y axis}\label{Appendix:MG_general}
In what follows we identify the main contributions to the term $\left\langle \Delta p^{\mathbf{B}}_{y,i} \right\rangle $ in Eq. \eqref{eq:offset_terms} for the recolliding and bound electrons. We find $ \Delta p^{\mathbf{B}}_{y,i} $ using a simple model of an electron inside an electromagnetic field and account for the effect of the Coulomb forces with a sharp change during recollision in the momentum of each electron $[\Delta p_{z,i} (t_{\mathrm{rec}})]$.
\subsubsection{Momentum change along the y axis for a recolliding electron}\label{appendix:MG_rec}
The Lorentz force acting on an electron $i$ is
\begin{equation}\label{eq:LF}
\mathbf{F}_{\mathrm{L}} = -\left[ \mathbf{E}(y_i,t) + \mathbf{p}_i\times \mathbf{B}(y_i,t) \right].
\end{equation}
The momentum of the electron $i$ at time $t$ is then given by
\begin{equation}\label{eq:momentum}
\mathbf{p}_i(t)= \mathbf{p}_i(t_0)-\int_{t_0}^{t} \left[ \mathbf{E}(y_i,t') + \mathbf{p}_i\times \mathbf{B}(y_i,t') \right] dt'.
\end{equation}
Eq. \eqref{eq:momentum} does not account for the Coulomb interaction between an electron and the core as well as between electrons. In a simplified model for the recolliding electron, we account for the momentum change due to a recollision and hence due to the Coulomb forces by adding a term in \eqref{eq:momentum} as follows
\begin{align}\label{eq:momentum2}
\begin{split}
\mathbf{p}_{i}(t)&=\mathbf{p}_{i}(t_0)-\int_{t_0}^{t}  \left[ \mathbf{E}(y_i,t') + \mathbf{p}_i \times \mathbf{B}(y_i,t') \right] dt' \\
&+ \mathrm{H}(t-t_{\mathrm{rec}})\Delta\mathbf{p}_{i}(t_{\mathrm{rec}}),
\end{split}
\end{align}
with $\mathrm{H}(t-t_{\mathrm{rec}})$ the Heaviside function \cite{abramowitz1965handbook} and with $\Delta\mathbf{p}_{i}(t_{\mathrm{rec}})$ being the momentum change due to the Coulomb forces just after and before the recollision time, $t_{rec}$. Then from Eq. \eqref{eq:momentum2} it follows that the $y$ component of the momentum change due to the magnetic field for $t>t_{\mathrm{rec}}$ takes the form
\begin{widetext}
\begin{align}\label{eq:momentumgaintotal}
\begin{split}
\Delta p_{y,i}^{\mathbf{B}}(t_0 \to t)&=  - \int_{t_0}^{t} p_{z,i}(t')B(y_i,t')dt'\\
&= -\int_{t_0}^{t} \left[ p_{z,i}(t_0) - \int_{t_0}^{t'} E(y_i,t'')dt'' + \int_{t_0}^{t'} p_{y,i}(t'')B(y_i,t'')dt'' + \mathrm{H}(t'-t_{\mathrm{rec}})\Delta p_{z,i}(t_{ \mathrm{rec} })\right] B(y_i,t')dt' \\ 
&= -\int_{t_0}^{t} \left[ p_{z,i}(t_0)  - \int_{t_0}^{t'} E(y_i,t'')dt'' + \int_{t_0}^{t'} p_{y,i}(t'')B(y_i,t'')dt'' \right] B(y_i,t')dt' - \Delta p_{z,i}(t_{ \mathrm{rec} })  \int_{t_{\mathrm{rec}}}^{t} B(y_i,t')dt'\\
&= \Delta p_{y,i}^{\mathbf{B},1}(t_0 \to t)-\Delta p_{z,i}(t_{ \mathrm{rec} })  \int_{t_{\mathrm{rec}}}^{t} B(y_i,t')dt'= \Delta p_{y,i}^{\mathbf{B},1}(t_0 \to t)+ \Delta p_{y,i}^{\mathbf{B},2}(t_{\mathrm{rec}} \to t)
\end{split}
\end{align}
\end{widetext}
 The term $\Delta p_{y,i}^{\mathbf{B},1}(t_0 \to t)$ simplifies when we take into account that in our model the initial  momentum of the recolliding electron along the direction of the electric field is zero, $ p_{z,i}(t_0)=0$. Furthermore, for the purposes of this model we neglect terms of the order of $\mathbf{B}^2$, since the ratio of the magnitudes of the electric and magnetic field is $| E(y_i,t)/B(y_i,t) | = c.$ Another approximation we make for the purposes of this model is that we compute the integral of the magnetic and electric field over time at the position $y_i=0$. That is,
 \begin{align}
E(y_i,t)&\approx E(0,t)\equiv E(t)\\
B(y_i,t)&\approx B(0,t)\equiv B(t).
 \end{align} 
Given the above approximations we find that 
\begin{subequations}\label{eq:momentumgains}
\begin{align}
\Delta p_{y,i}^{\mathbf{B},1}(t_0 \to t) &= \int_{t_0}^{t} \left[ \int_{t_0}^{t'} E(t'')dt'' \right] B(t')dt' \label{eq:momentumgains_a}\\
\Delta p_{y,i}^{\mathbf{B},2}(t_{\mathrm{rec}} \to t) &= - \Delta p_{z,i}(t_{ \mathrm{rec} })  \int_{t_{\mathrm{rec}}}^{t} B(t')dt'.\label{eq:momentumgains_b}
\end{align}
\end{subequations}
 
For the recolliding electron,  at all three intensities both for TI and DI, we find that the term $\Delta p_{y,i}^{\mathbf{B},2}(t_{\mathrm{rec}} \to t_f)$ contributes the most to $\Delta p_{y,i}^{\mathbf{B}}(t_{0} \to t_f)$. To do so, we obtain $\Delta p_{z,i}(t_{\mathrm{rec}})$ from our full calculations using the ECBB model. Next, we show that $\Delta p_{y,i}^{\mathbf{B},2}(t_{\mathrm{rec}} \to t_f)$ is always positive. Indeed, we rewrite Eq. \eqref{eq:momentumgains_b} as

\begin{align}\label{eq:Appendix_MG_inc_v2}
\begin{split}
\Delta p_{y,i}^{\mathbf{B},2}(t_{\mathrm{rec}} \to t_f) &=-\Delta p_{z,i} (t_{ \mathrm{rec} }) \int_{t_{\mathrm{rec}}}^{t_f} \dfrac{E(t)}{c}dt \\
&= -\frac{1}{c} \Delta p_{z,i} (t_{ \mathrm{rec} }) \left[ A(t_{\mathrm{rec}}) - A(t_{f}) \right]  \\
&= -\frac{1}{c} \Delta p_{z,i} (t_{ \mathrm{rec} }) A(t_{\mathrm{rec}}), 
\end{split}
\end{align}
 where we use $\mathbf{E}(t) = -\frac{\partial\mathbf{A}(t)}{\partial t}$ and $A(t_{f}\to\infty)=0.$ Moreover, for the tunnelling/recolliding electron, we find that 
 \begin{align}
 \begin{split}
 p_{z,i}(t_{\mathrm{rec}})&= - \int_{t_0}^{t_{\mathrm{rec}}} E(t)dt \\
 &= - \left[  A(t_{0})  - A(t_{\mathrm{rec}}) \right] \\ 
 &= A(t_{\mathrm{rec}}),
 \end{split}
 \end{align}
 where we have used that $ A(t_{0}) \approx 0,$ since the electron tunnels at an initial time $t_0$ around an extremum of the electric field. Then Eq. \eqref{eq:Appendix_MG_inc_v2} can be written as
\begin{align}\label{eq:Appendix_MG_inc_v3} 
\begin{split}
&\Delta p_{y,i}^{\mathbf{B},2}(t_{\mathrm{rec}} \to t_f) = -\frac{1}{c} \Delta p_{z,i}(t_{ \mathrm{rec} }) p_{z,i}(t_{\mathrm{rec}})\\
&=-\frac{1}{c} \left[  p_{z,i}(t_{\mathrm{rec}} + \Delta t)-p_{z,i}(t_{\mathrm{rec}}) \right] p_{z,i}(t_{\mathrm{rec}}).
\end{split}
\end{align}
During a recollision, the magnitude of the momentum of the recolliding electron after the recollision is always smaller than its magnitude before the recollision, i.e. 
\begin{equation}\label{eq:Appendix_MG_inc_v4} 
|  p_{z,i}(t_{\mathrm{rec}} + \Delta t)-p_{z,i}(t_{\mathrm{rec}})  |<|p_{z,i}(t_{\mathrm{rec}})|.
\end{equation}
Combining Eqs. \eqref{eq:Appendix_MG_inc_v3} and \eqref{eq:Appendix_MG_inc_v4}, it is easy to show that $\Delta p_{y,i}^{\mathbf{B},2}(t_{\mathrm{rec}} \to t_f)$ is always greater than zero.

\subsubsection{Momentum change along the y axis for a bound electron}\label{appendix:MG_bound}
Concerning a bound electron, we assume that the electron feels the electric and magnetic field only after it is ionized, i.e. roughly at the recollision time. Hence, for the bound electron, in \eqref{eq:momentum2} and \eqref{eq:momentumgaintotal} we substitute $t_0$ by $t_{\mathrm{rec}}$. We also assume that $p_{i}(t_{\mathrm{rec}}) \approx 0$. Given the above, we find that 

\begin{align}\label{eq:momentumgaintotalBound}
\begin{split}
&\Delta p_{y,i}^{\mathbf{B}}(t_{rec} \to t)= \int_{t_{\mathrm{rec}}}^{t} \left[ \int_{t_{\mathrm{rec}}}^{t'} E(t'')dt'' \right] B(t')dt'\\
&\hspace{3cm} -\Delta p_{z,i}(t_{\mathrm{rec}} )  \int_{t_{\mathrm{rec}}}^{t} B(t')dt'\\ 
&= \Delta p_{y,i}^{\mathbf{B},1}(t_{\mathrm{rec}} \to t)+ \Delta p_{y,i}^{\mathbf{B},2}(t_{\mathrm{rec}} \to t)
\end{split}
\end{align}
For both TI and DI, at all three intensities, we find that the term $\Delta p_{y,i}^{\mathbf{B},1}(t_{\mathrm{rec}} \to t_f)$ contributes the most to $\Delta p_{y,i}^{\mathbf{B}}(t_{\mathrm{rec}} \to t_f)$ for the bound electron. Next, we show that this term is always positive as follows 
\begin{align}\label{eq:momentumgainsBound_a}
\begin{split}
&\Delta p_{y,i}^{\mathbf{B},1}(t_{\mathrm{rec}} \to t_f) = \int_{t_{\mathrm{rec}}}^{t_f} \left[ \int_{t_{\mathrm{rec}}}^{t} E(t')dt' \right] B(t)dt \\
&= \int_{t_{\mathrm{rec}}}^{t_f} \left[ A(t_{\mathrm{rec}})-A(t)\right] B(t)dt  \\
&=\left[ A(t_{\mathrm{rec}})\int_{t_{\mathrm{rec}}}^{t_f}B(t)dt  - \int_{t_{\mathrm{rec}}}^{t_f}A(t) B(t)dt  \right]\\
&= \left\{\dfrac{A(t_{\mathrm{rec}})}{c} \left[A(t_{\mathrm{rec}})-A(t_f) \right]  - \int_{t_{\mathrm{rec}}}^{t_f}A(t)\dfrac{E(t)}{c}dt \right\}\\
&=\frac{1}{c}\left[ A^2(t_{\mathrm{rec}}) + \dfrac{A^2(t_f)}{2} - \dfrac{ A^2(t_{\mathrm{rec}}) }{2}\right]\\
&=\frac{1}{2c} A^2(t_{\mathrm{rec}}) > 0.
\end{split}
\end{align}
where we use $A(t_{f}\to\infty)=0.$

\subsection{Comparison of the offset between DI and TI}
\begin{figure}[t]
\centering
\includegraphics[width=\linewidth]{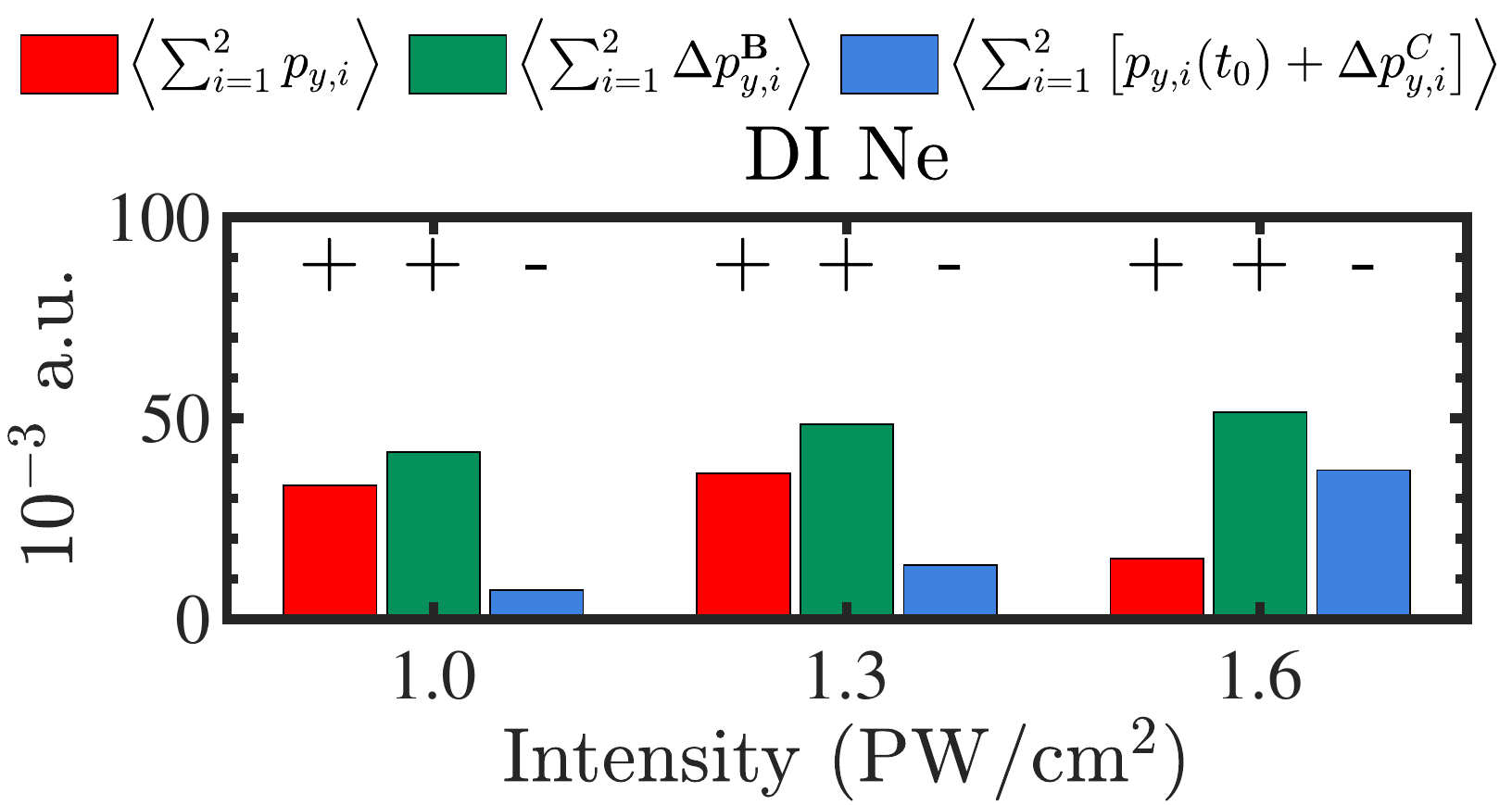}
\caption{For Ne, at each intensity, height of red bar denotes the momentum offset per pair of electrons for DI $\left\langle \sum_{i=1}^{2} p_{y,i} \right\rangle$, and the contributions due to the magnetic field $\left\langle \sum_{i=1}^{2} \Delta p_{y,i}^{\mathbf{B}} \right\rangle$ (green bar) and due to the initial momentum and Coulomb forces $\left\langle \sum_{i=1}^{2} \left[ p_{y,i}(t_0)+\Delta p_{y,i}^{C} \right]\right\rangle$ (blue bar). The plus (+), minus (-) sign above the bar denotes a positive or negative value, respectively, for the given contribution.}\label{fig:Offset_all_events_DI}
\end{figure}

In \fig{fig:Offset_all_events_DI}, for DI of driven Ne, we compute the $y$ component of the average sum of the final momenta of the ionizing electron pair, $\left\langle \sum_{i=1}^{2} p_{y,i} \right\rangle_{\text{DI}}$. This momentum offset is denoted by the height of the red bar. At intensities 1.0 and 1.3 $\mathrm{PW/cm^2}$ we find that the momentum offset has a positive value around 0.035 a.u. This is roughly three times larger than twice (to account for the electron pair) the momentum offset in single ionization. At intensity 1.6 $\mathrm{PW/cm^2}$ the value of the positive momentum offset is approximately half its value at the two smaller intensities. \fig{fig:Offset_all_events_DI}, clearly shows that the positive momentum offset for DI is due to the momentum change  from the magnetic field (green bars), as was the case for TI. At all three intensities, we find that for triple ionization $2/3\left\langle \sum_{i=1}^{3} \Delta p_{y,i}^{\mathbf{B}} \right\rangle$ ranges roughly between 0.06 to 0.07 a.u. (green bars in \fig{fig:Offset_all_events_TI}), while for double ionization $\left\langle \sum_{i=1}^{2} \Delta p_{y,i}^{\mathbf{B}} \right\rangle$ is smaller, ranging roughly between 0.04 to 0.05 a.u.  (green bars in \fig{fig:Offset_all_events_DI}). 

Now, we show that the smaller positive momentum offset due to the magnetic field in double ionization compared to triple ionization is consistent with the simple model developed in Section \ref{Appendix:MG_general}. Indeed, recollisions are stronger in TI versus DI. This is evidenced by the higher degree of electron-electron correlation in TI compared to DI, compare top with bottom row in \fig{fig:Correlated_momenta}. A stronger recollision in TI translates to a larger change of the $z$ component of the momentum of the recolliding electron due to the Coulomb forces during recollision, i.e., to a larger value of $\Delta p_{z,i}(t_{\mathrm{rec}})$ in Eq. \eqref{eq:Appendix_MG_inc_v2}. Moreover, a stronger recollision also translates to the time of recollision being around a zero of the electric field, resulting to an extremum of $A(t_{\mathrm{rec}})$. Hence, the most important contributions to the momentum offset, for the recolliding electron the term $\Delta p_{y,i}^{\mathbf{B},2}(t_{\mathrm{rec}} \to t_f)$ [Eq. \eqref{eq:Appendix_MG_inc_v2}] and for the bound electron the term $\Delta p_{y,i}^{\mathbf{B},1}(t_{\mathrm{rec}} \to t_f)$ [Eq. \eqref{eq:momentumgainsBound_a}], have larger values for TI compared to DI.

\subsection{Momentum offset for direct versus delayed pathways in TI and DI}
\begin{figure}[b]
\centering
\includegraphics[width=\linewidth]{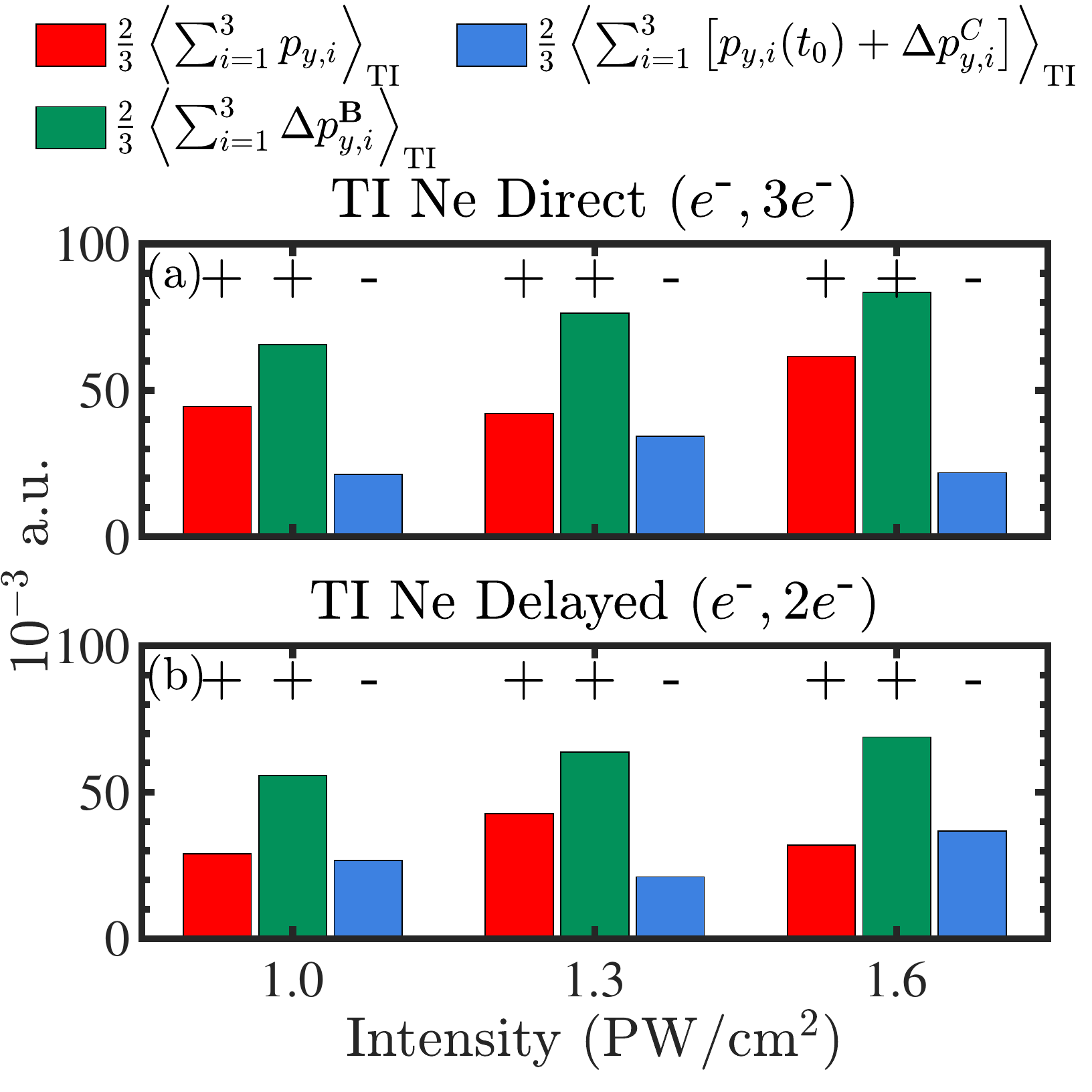}
\caption{For Ne, at each intensity, height of red bar denotes the momentum offset per pair of electrons for TI $2/3\left\langle \sum_{i=1}^{3} p_{y,i} \right\rangle$, and the contributions due to the magnetic field $2/3\left\langle \sum_{i=1}^{3} \Delta p_{y,i}^{\mathbf{B}} \right\rangle$ (green bar) and due to the initial momentum and Coulomb forces $2/3\left\langle \sum_{i=1}^{3} \left[ p_{y,i}(t_0)+\Delta p_{y,i}^{C} \right]\right\rangle$ (blue bar). The plus (+), minus (-) sign above the bar denotes a positive or negative value, respectively, for the given contribution. The top row corresponds to the direct $(e^{\text{-}},3e^\text{-})$ pathway and the bottom row to the delayed $(e^{\text{-}},2e^\text{-})$ pathway.}\label{fig:Offset_pathways_TI}
\end{figure}
\begin{figure}[t]
\centering
\includegraphics[width=\linewidth]{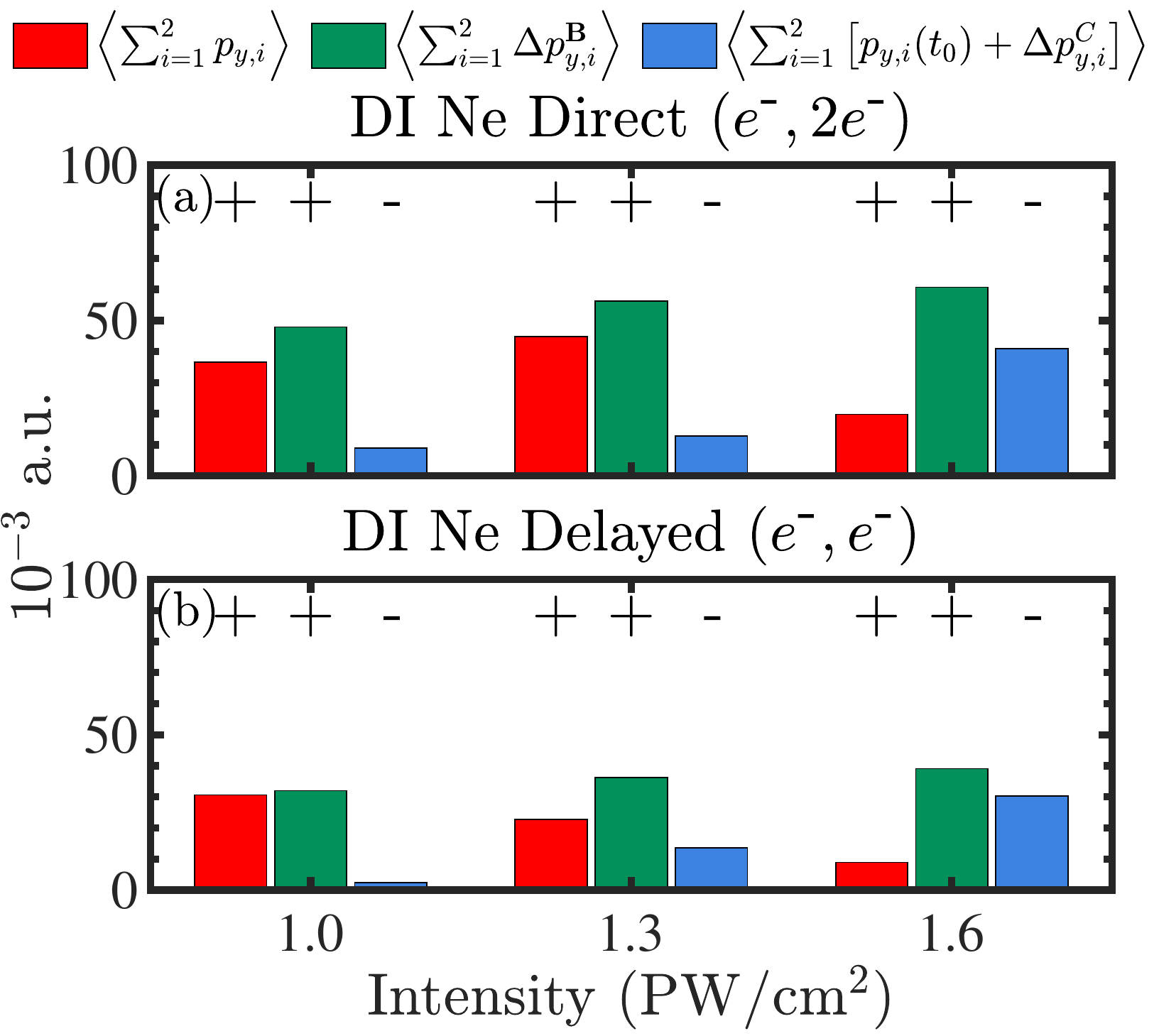}
\caption{For Ne, at each intensity, height of red bar denotes the momentum offset per pair of electrons for DI $\left\langle \sum_{i=1}^{2} p_{y,i} \right\rangle$, and the contributions due to the magnetic field $\left\langle \sum_{i=1}^{2} \Delta p_{y,i}^{\mathbf{B}} \right\rangle$ (green bar) and due to the initial momentum and Coulomb forces $\left\langle \sum_{i=1}^{2} \left[ p_{y,i}(t_0)+\Delta p_{y,i}^{C} \right]\right\rangle$ (blue bar). The plus (+), minus (-) sign above the bar denotes a positive or negative value, respectively, for the given contribution. The top row corresponds to the direct $(e^{\text{-}},2e^{\text{-}})$ pathway and the bottom row to the delayed $(e^{\text{-}},e^{\text{-}}$) pathway.}\label{fig:Offset_pathways_DI}
\end{figure}

In what follows, we compare the momentum offset in the direct versus the delayed pathway both in TI and DI. 
In \fig{fig:Offset_pathways_TI}, for TI of driven Ne, we show the momentum offset (red bars), the contribution to this offset from the magnetic field (green bars) as well as the contribution from the Coulomb forces (blue bars) for the direct (top row) and the delayed $(e^{\text{-}},2e^\text{-})$ pathway (bottom row). At all three intensities we find that the momentum offset (red bars) is larger in the direct compared to the delayed pathway. \fig{fig:Offset_pathways_TI} clearly shows that this is mainly due to the larger positive values of the momentum change due to the magnetic field (green bars) in the direct compared to the delayed pathway. That is, the term $2/3\left\langle \sum_{i=1}^{3} \Delta p_{y,i}^{\mathbf{B}} \right\rangle$ is larger in the direct compared to the delayed pathway. \fig{fig:Offset_pathways_DI}. shows that the same holds true for DI of driven Ne. That is, the momentum offset as well as the contribution to this offset from the magnetic field, $\left\langle \sum_{i=1}^{2} \Delta p_{y,i}^{\mathbf{B}} \right\rangle$, is larger in the direct compared to the delayed pathway. Next, we explain why this is the case. During recollision, the recolliding electron gives more energy to the bound electrons in the direct compared to the delayed pathway. That is, the sharp momentum change of the recolliding electron during recollision, $\Delta p_{z,i}(t_{\mathrm{rec}})$, is larger in the direct compared to the delayed pathway. Hence, $\Delta p_{y,i}^{\mathbf{B},2}(t_{\mathrm{rec}} \to t_f)$ in Eq. \eqref{eq:Appendix_MG_inc_v2} for the recolliding electron, is larger in the direct pathway. In addition, for the bound electrons, $\Delta p_{y,i}^{\mathbf{B},1}(t_{\mathrm{rec}} \to t_f)$ in Eq. \eqref{eq:momentumgainsBound_a} is larger in the direct compared to the delayed pathway. The reason is that both bound electrons in the direct pathway ionize soon after the recollision time which is around an extremum of the vector potential $A$, i.e., maximum value of $A(t_{\mathrm{rec}})$. However, in the delayed pathway, most likely, it is one of the bound electrons that ionizes with a delay from the recollision time and hence $A(t_{\mathrm{rec}})$ is smaller than its extremum value.

\section{Conclusions}
In conclusion, we use the ECBB 3D semiclassical model to identify nondipole effects in triple and double ionization in Ne driven by infrared pulses for intensities where recollisions, i.e., electron-electron correlation, prevail. We find a large positive average sum of the final electron momenta along the direction of light propagation. This momentum offset is zero in the absence of the magnetic field. Most importantly, we show this final electron momentum offset to be a probe of electron-electron correlation. Indeed,  we find a larger momentum offset for the more correlated electron-electron ionization  i) in triple compared to double ionization of driven Ne, especially at high intensities, and ii) in  the direct versus the delayed pathway of triple  and double ionization of Ne. The nondipole effects identified here in  multielectron ionization observables  can  be accessed  and hence verified by future experiments.

\begin{acknowledgements}
A.E. and G.P.K. acknowledge the EPSRC Grant EP/W005352/1. The authors acknowledge the use of the UCL Myriad High Performance Computing Facility (Myriad@UCL), the use of the UCL Kathleen High Performance Computing Facility (Kathleen@UCL), and associated support services, in the completion of this work.
\end{acknowledgements}

\bibliographystyle{apsrev4-1}
\bibliography{bibliography}

\begin{thebibliography}{42}%
\makeatletter
\providecommand \@ifxundefined [1]{%
 \@ifx{#1\undefined}
}%
\providecommand \@ifnum [1]{%
 \ifnum #1\expandafter \@firstoftwo
 \else \expandafter \@secondoftwo
 \fi
}%
\providecommand \@ifx [1]{%
 \ifx #1\expandafter \@firstoftwo
 \else \expandafter \@secondoftwo
 \fi
}%
\providecommand \natexlab [1]{#1}%
\providecommand \enquote  [1]{``#1''}%
\providecommand \bibnamefont  [1]{#1}%
\providecommand \bibfnamefont [1]{#1}%
\providecommand \citenamefont [1]{#1}%
\providecommand \href@noop [0]{\@secondoftwo}%
\providecommand \href [0]{\begingroup \@sanitize@url \@href}%
\providecommand \@href[1]{\@@startlink{#1}\@@href}%
\providecommand \@@href[1]{\endgroup#1\@@endlink}%
\providecommand \@sanitize@url [0]{\catcode `\\12\catcode `\$12\catcode
  `\&12\catcode `\#12\catcode `\^12\catcode `\_12\catcode `\%12\relax}%
\providecommand \@@startlink[1]{}%
\providecommand \@@endlink[0]{}%
\providecommand \url  [0]{\begingroup\@sanitize@url \@url }%
\providecommand \@url [1]{\endgroup\@href {#1}{\urlprefix }}%
\providecommand \urlprefix  [0]{URL }%
\providecommand \Eprint [0]{\href }%
\providecommand \doibase [0]{http://dx.doi.org/}%
\providecommand \selectlanguage [0]{\@gobble}%
\providecommand \bibinfo  [0]{\@secondoftwo}%
\providecommand \bibfield  [0]{\@secondoftwo}%
\providecommand \translation [1]{[#1]}%
\providecommand \BibitemOpen [0]{}%
\providecommand \bibitemStop [0]{}%
\providecommand \bibitemNoStop [0]{.\EOS\space}%
\providecommand \EOS [0]{\spacefactor3000\relax}%
\providecommand \BibitemShut  [1]{\csname bibitem#1\endcsname}%
\let\auto@bib@innerbib\@empty
\bibitem [{\citenamefont {l'Huillier}\ \emph {et~al.}(1983)\citenamefont
  {l'Huillier}, \citenamefont {Lompre}, \citenamefont {Mainfray},\ and\
  \citenamefont {Manus}}]{knee1983}%
  \BibitemOpen
  \bibfield  {author} {\bibinfo {author} {\bibfnamefont {A.}~\bibnamefont
  {l'Huillier}}, \bibinfo {author} {\bibfnamefont {L.~A.}\ \bibnamefont
  {Lompre}}, \bibinfo {author} {\bibfnamefont {G.}~\bibnamefont {Mainfray}}, \
  and\ \bibinfo {author} {\bibfnamefont {C.}~\bibnamefont {Manus}},\ }\href
  {\doibase 10.1103/PhysRevA.27.2503} {\bibfield  {journal} {\bibinfo
  {journal} {Phys. Rev. A}\ }\textbf {\bibinfo {volume} {27}},\ \bibinfo
  {pages} {2503} (\bibinfo {year} {1983})}\BibitemShut {NoStop}%
\bibitem [{\citenamefont {Emelin}\ and\ \citenamefont
  {Ryabikin}(2014)}]{stabilization}%
  \BibitemOpen
  \bibfield  {author} {\bibinfo {author} {\bibfnamefont {M.~Y.}\ \bibnamefont
  {Emelin}}\ and\ \bibinfo {author} {\bibfnamefont {M.~Y.}\ \bibnamefont
  {Ryabikin}},\ }\href {\doibase 10.1103/PhysRevA.89.013418} {\bibfield
  {journal} {\bibinfo  {journal} {Phys. Rev. A}\ }\textbf {\bibinfo {volume}
  {89}},\ \bibinfo {pages} {013418} (\bibinfo {year} {2014})}\BibitemShut
  {NoStop}%
\bibitem [{\citenamefont {Chiril\ifmmode~\u{a}\else \u{a}\fi{}}\ \emph
  {et~al.}(2002)\citenamefont {Chiril\ifmmode~\u{a}\else \u{a}\fi{}},
  \citenamefont {Kylstra}, \citenamefont {Potvliege},\ and\ \citenamefont
  {Joachain}}]{HHG1}%
  \BibitemOpen
  \bibfield  {author} {\bibinfo {author} {\bibfnamefont {C.~C.}\ \bibnamefont
  {Chiril\ifmmode~\u{a}\else \u{a}\fi{}}}, \bibinfo {author} {\bibfnamefont
  {N.~J.}\ \bibnamefont {Kylstra}}, \bibinfo {author} {\bibfnamefont {R.~M.}\
  \bibnamefont {Potvliege}}, \ and\ \bibinfo {author} {\bibfnamefont {C.~J.}\
  \bibnamefont {Joachain}},\ }\href {\doibase 10.1103/PhysRevA.66.063411}
  {\bibfield  {journal} {\bibinfo  {journal} {Phys. Rev. A}\ }\textbf {\bibinfo
  {volume} {66}},\ \bibinfo {pages} {063411} (\bibinfo {year}
  {2002})}\BibitemShut {NoStop}%
\bibitem [{\citenamefont {Walser}\ \emph {et~al.}(2000)\citenamefont {Walser},
  \citenamefont {Keitel}, \citenamefont {Scrinzi},\ and\ \citenamefont
  {Brabec}}]{HHG2}%
  \BibitemOpen
  \bibfield  {author} {\bibinfo {author} {\bibfnamefont {M.~W.}\ \bibnamefont
  {Walser}}, \bibinfo {author} {\bibfnamefont {C.~H.}\ \bibnamefont {Keitel}},
  \bibinfo {author} {\bibfnamefont {A.}~\bibnamefont {Scrinzi}}, \ and\
  \bibinfo {author} {\bibfnamefont {T.}~\bibnamefont {Brabec}},\ }\href
  {\doibase 10.1103/PhysRevLett.85.5082} {\bibfield  {journal} {\bibinfo
  {journal} {Phys. Rev. Lett.}\ }\textbf {\bibinfo {volume} {85}},\ \bibinfo
  {pages} {5082} (\bibinfo {year} {2000})}\BibitemShut {NoStop}%
\bibitem [{\citenamefont {Keitel}\ and\ \citenamefont {Knight}(1995)}]{HHG3}%
  \BibitemOpen
  \bibfield  {author} {\bibinfo {author} {\bibfnamefont {C.~H.}\ \bibnamefont
  {Keitel}}\ and\ \bibinfo {author} {\bibfnamefont {P.~L.}\ \bibnamefont
  {Knight}},\ }\href {\doibase 10.1103/PhysRevA.51.1420} {\bibfield  {journal}
  {\bibinfo  {journal} {Phys. Rev. A}\ }\textbf {\bibinfo {volume} {51}},\
  \bibinfo {pages} {1420} (\bibinfo {year} {1995})}\BibitemShut {NoStop}%
\bibitem [{\citenamefont {Palaniyappan}\ \emph {et~al.}(2005)\citenamefont
  {Palaniyappan}, \citenamefont {DiChiara}, \citenamefont {Chowdhury},
  \citenamefont {Falkowski}, \citenamefont {Ongadi}, \citenamefont {Huskins},\
  and\ \citenamefont {Walker}}]{Neon}%
  \BibitemOpen
  \bibfield  {author} {\bibinfo {author} {\bibfnamefont {S.}~\bibnamefont
  {Palaniyappan}}, \bibinfo {author} {\bibfnamefont {A.}~\bibnamefont
  {DiChiara}}, \bibinfo {author} {\bibfnamefont {E.}~\bibnamefont {Chowdhury}},
  \bibinfo {author} {\bibfnamefont {A.}~\bibnamefont {Falkowski}}, \bibinfo
  {author} {\bibfnamefont {G.}~\bibnamefont {Ongadi}}, \bibinfo {author}
  {\bibfnamefont {E.~L.}\ \bibnamefont {Huskins}}, \ and\ \bibinfo {author}
  {\bibfnamefont {B.~C.}\ \bibnamefont {Walker}},\ }\href {\doibase
  10.1103/PhysRevLett.94.243003} {\bibfield  {journal} {\bibinfo  {journal}
  {Phys. Rev. Lett.}\ }\textbf {\bibinfo {volume} {94}},\ \bibinfo {pages}
  {243003} (\bibinfo {year} {2005})}\BibitemShut {NoStop}%
\bibitem [{\citenamefont {Reiss}(2008)}]{Magnetic1}%
  \BibitemOpen
  \bibfield  {author} {\bibinfo {author} {\bibfnamefont {H.~R.}\ \bibnamefont
  {Reiss}},\ }\href {\doibase 10.1103/PhysRevLett.101.043002} {\bibfield
  {journal} {\bibinfo  {journal} {Phys. Rev. Lett.}\ }\textbf {\bibinfo
  {volume} {101}},\ \bibinfo {pages} {043002} (\bibinfo {year}
  {2008})}\BibitemShut {NoStop}%
\bibitem [{\citenamefont {Reiss}(2014)}]{Magnetic2}%
  \BibitemOpen
  \bibfield  {author} {\bibinfo {author} {\bibfnamefont {H.~R.}\ \bibnamefont
  {Reiss}},\ }\href {\doibase 10.1088/0953-4075/47/20/204006} {\bibfield
  {journal} {\bibinfo  {journal} {J. Phys. B}\ }\textbf {\bibinfo {volume}
  {47}},\ \bibinfo {pages} {204006} (\bibinfo {year} {2014})}\BibitemShut
  {NoStop}%
\bibitem [{\citenamefont {Smeenk}\ \emph {et~al.}(2011)\citenamefont {Smeenk},
  \citenamefont {Arissian}, \citenamefont {Zhou}, \citenamefont {Mysyrowicz},
  \citenamefont {Villeneuve}, \citenamefont {Staudte},\ and\ \citenamefont
  {Corkum}}]{PhysRevLett.106.193002}%
  \BibitemOpen
  \bibfield  {author} {\bibinfo {author} {\bibfnamefont {C.~T.~L.}\
  \bibnamefont {Smeenk}}, \bibinfo {author} {\bibfnamefont {L.}~\bibnamefont
  {Arissian}}, \bibinfo {author} {\bibfnamefont {B.}~\bibnamefont {Zhou}},
  \bibinfo {author} {\bibfnamefont {A.}~\bibnamefont {Mysyrowicz}}, \bibinfo
  {author} {\bibfnamefont {D.~M.}\ \bibnamefont {Villeneuve}}, \bibinfo
  {author} {\bibfnamefont {A.}~\bibnamefont {Staudte}}, \ and\ \bibinfo
  {author} {\bibfnamefont {P.~B.}\ \bibnamefont {Corkum}},\ }\href {\doibase
  10.1103/PhysRevLett.106.193002} {\bibfield  {journal} {\bibinfo  {journal}
  {Phys. Rev. Lett.}\ }\textbf {\bibinfo {volume} {106}},\ \bibinfo {pages}
  {193002} (\bibinfo {year} {2011})}\BibitemShut {NoStop}%
\bibitem [{\citenamefont {Ludwig}\ \emph {et~al.}(2014)\citenamefont {Ludwig},
  \citenamefont {Maurer}, \citenamefont {Mayer}, \citenamefont {Phillips},
  \citenamefont {Gallmann},\ and\ \citenamefont {Keller}}]{KellerMagnetic2014}%
  \BibitemOpen
  \bibfield  {author} {\bibinfo {author} {\bibfnamefont {A.}~\bibnamefont
  {Ludwig}}, \bibinfo {author} {\bibfnamefont {J.}~\bibnamefont {Maurer}},
  \bibinfo {author} {\bibfnamefont {B.~W.}\ \bibnamefont {Mayer}}, \bibinfo
  {author} {\bibfnamefont {C.~R.}\ \bibnamefont {Phillips}}, \bibinfo {author}
  {\bibfnamefont {L.}~\bibnamefont {Gallmann}}, \ and\ \bibinfo {author}
  {\bibfnamefont {U.}~\bibnamefont {Keller}},\ }\href {\doibase
  10.1103/PhysRevLett.113.243001} {\bibfield  {journal} {\bibinfo  {journal}
  {Phys. Rev. Lett.}\ }\textbf {\bibinfo {volume} {113}},\ \bibinfo {pages}
  {243001} (\bibinfo {year} {2014})}\BibitemShut {NoStop}%
\bibitem [{\citenamefont {Chelkowski}\ \emph {et~al.}(2014)\citenamefont
  {Chelkowski}, \citenamefont {Bandrauk},\ and\ \citenamefont
  {Corkum}}]{CorkumBandrauk2015}%
  \BibitemOpen
  \bibfield  {author} {\bibinfo {author} {\bibfnamefont {S.}~\bibnamefont
  {Chelkowski}}, \bibinfo {author} {\bibfnamefont {A.~D.}\ \bibnamefont
  {Bandrauk}}, \ and\ \bibinfo {author} {\bibfnamefont {P.~B.}\ \bibnamefont
  {Corkum}},\ }\href {\doibase 10.1103/PhysRevLett.113.263005} {\bibfield
  {journal} {\bibinfo  {journal} {Phys. Rev. Lett.}\ }\textbf {\bibinfo
  {volume} {113}},\ \bibinfo {pages} {263005} (\bibinfo {year}
  {2014})}\BibitemShut {NoStop}%
\bibitem [{\citenamefont {Wolter}\ \emph {et~al.}(2015)\citenamefont {Wolter},
  \citenamefont {Pullen}, \citenamefont {Baudisch}, \citenamefont {Sclafani},
  \citenamefont {Hemmer}, \citenamefont {Senftleben}, \citenamefont
  {Schr\"oter}, \citenamefont {Ullrich}, \citenamefont {Moshammer},\ and\
  \citenamefont {Biegert}}]{Biegert2015}%
  \BibitemOpen
  \bibfield  {author} {\bibinfo {author} {\bibfnamefont {B.}~\bibnamefont
  {Wolter}}, \bibinfo {author} {\bibfnamefont {M.~G.}\ \bibnamefont {Pullen}},
  \bibinfo {author} {\bibfnamefont {M.}~\bibnamefont {Baudisch}}, \bibinfo
  {author} {\bibfnamefont {M.}~\bibnamefont {Sclafani}}, \bibinfo {author}
  {\bibfnamefont {M.}~\bibnamefont {Hemmer}}, \bibinfo {author} {\bibfnamefont
  {A.}~\bibnamefont {Senftleben}}, \bibinfo {author} {\bibfnamefont {C.~D.}\
  \bibnamefont {Schr\"oter}}, \bibinfo {author} {\bibfnamefont
  {J.}~\bibnamefont {Ullrich}}, \bibinfo {author} {\bibfnamefont
  {R.}~\bibnamefont {Moshammer}}, \ and\ \bibinfo {author} {\bibfnamefont
  {J.}~\bibnamefont {Biegert}},\ }\href {\doibase 10.1103/PhysRevX.5.021034}
  {\bibfield  {journal} {\bibinfo  {journal} {Phys. Rev. X}\ }\textbf {\bibinfo
  {volume} {5}},\ \bibinfo {pages} {021034} (\bibinfo {year}
  {2015})}\BibitemShut {NoStop}%
\bibitem [{\citenamefont {Emmanouilidou}\ and\ \citenamefont
  {Meltzer}(2017)}]{Emmanouilidou1}%
  \BibitemOpen
  \bibfield  {author} {\bibinfo {author} {\bibfnamefont {A.}~\bibnamefont
  {Emmanouilidou}}\ and\ \bibinfo {author} {\bibfnamefont {T.}~\bibnamefont
  {Meltzer}},\ }\href {\doibase 10.1103/PhysRevA.95.033405} {\bibfield
  {journal} {\bibinfo  {journal} {Phys. Rev. A}\ }\textbf {\bibinfo {volume}
  {95}},\ \bibinfo {pages} {033405} (\bibinfo {year} {2017})}\BibitemShut
  {NoStop}%
\bibitem [{\citenamefont {Emmanouilidou}\ \emph {et~al.}(2017)\citenamefont
  {Emmanouilidou}, \citenamefont {Meltzer},\ and\ \citenamefont
  {Corkum}}]{Emmanouilidou2}%
  \BibitemOpen
  \bibfield  {author} {\bibinfo {author} {\bibfnamefont {A.}~\bibnamefont
  {Emmanouilidou}}, \bibinfo {author} {\bibfnamefont {T.}~\bibnamefont
  {Meltzer}}, \ and\ \bibinfo {author} {\bibfnamefont {P.~B.}\ \bibnamefont
  {Corkum}},\ }\href {\doibase 10.1088/1361-6455/aa90e9} {\bibfield  {journal}
  {\bibinfo  {journal} {J. Phys. B}\ }\textbf {\bibinfo {volume} {50}},\
  \bibinfo {pages} {225602} (\bibinfo {year} {2017})}\BibitemShut {NoStop}%
\bibitem [{\citenamefont {Willenberg}\ \emph {et~al.}(2019)\citenamefont
  {Willenberg}, \citenamefont {Maurer}, \citenamefont {Mayer},\ and\
  \citenamefont {Keller}}]{KellerMagneticFields}%
  \BibitemOpen
  \bibfield  {author} {\bibinfo {author} {\bibfnamefont {B.}~\bibnamefont
  {Willenberg}}, \bibinfo {author} {\bibfnamefont {J.}~\bibnamefont {Maurer}},
  \bibinfo {author} {\bibfnamefont {B.~W.}\ \bibnamefont {Mayer}}, \ and\
  \bibinfo {author} {\bibfnamefont {U.}~\bibnamefont {Keller}},\ }\href
  {\doibase 10.1038/s41467-019-13409-6} {\bibfield  {journal} {\bibinfo
  {journal} {Nat. Commun.}\ }\textbf {\bibinfo {volume} {10}},\ \bibinfo
  {pages} {5548} (\bibinfo {year} {2019})}\BibitemShut {NoStop}%
\bibitem [{\citenamefont {Sun}\ \emph {et~al.}(2020)\citenamefont {Sun},
  \citenamefont {Chen}, \citenamefont {Zhang}, \citenamefont {Qiang},
  \citenamefont {Li}, \citenamefont {Lu}, \citenamefont {Gong}, \citenamefont
  {Ji}, \citenamefont {Lin}, \citenamefont {Li}, \citenamefont {Tong},
  \citenamefont {Chen}, \citenamefont {Ruiz}, \citenamefont {Wu},\ and\
  \citenamefont {He}}]{FSun}%
  \BibitemOpen
  \bibfield  {author} {\bibinfo {author} {\bibfnamefont {F.}~\bibnamefont
  {Sun}}, \bibinfo {author} {\bibfnamefont {X.}~\bibnamefont {Chen}}, \bibinfo
  {author} {\bibfnamefont {W.}~\bibnamefont {Zhang}}, \bibinfo {author}
  {\bibfnamefont {J.}~\bibnamefont {Qiang}}, \bibinfo {author} {\bibfnamefont
  {H.}~\bibnamefont {Li}}, \bibinfo {author} {\bibfnamefont {P.}~\bibnamefont
  {Lu}}, \bibinfo {author} {\bibfnamefont {X.}~\bibnamefont {Gong}}, \bibinfo
  {author} {\bibfnamefont {Q.}~\bibnamefont {Ji}}, \bibinfo {author}
  {\bibfnamefont {K.}~\bibnamefont {Lin}}, \bibinfo {author} {\bibfnamefont
  {H.}~\bibnamefont {Li}}, \bibinfo {author} {\bibfnamefont {J.}~\bibnamefont
  {Tong}}, \bibinfo {author} {\bibfnamefont {F.}~\bibnamefont {Chen}}, \bibinfo
  {author} {\bibfnamefont {C.}~\bibnamefont {Ruiz}}, \bibinfo {author}
  {\bibfnamefont {J.}~\bibnamefont {Wu}}, \ and\ \bibinfo {author}
  {\bibfnamefont {F.}~\bibnamefont {He}},\ }\href {\doibase
  10.1103/PhysRevA.101.021402} {\bibfield  {journal} {\bibinfo  {journal}
  {Phys. Rev. A}\ }\textbf {\bibinfo {volume} {101}},\ \bibinfo {pages}
  {021402(R)} (\bibinfo {year} {2020})}\BibitemShut {NoStop}%
\bibitem [{\citenamefont {Lin}\ \emph {et~al.}(2022{\natexlab{a}})\citenamefont
  {Lin}, \citenamefont {Brennecke}, \citenamefont {Ni}, \citenamefont {Chen},
  \citenamefont {Hartung}, \citenamefont {Trabert}, \citenamefont {Fehre},
  \citenamefont {Rist}, \citenamefont {Tong}, \citenamefont {Burgd\"orfer},
  \citenamefont {Schmidt}, \citenamefont {Sch\"offler}, \citenamefont {Jahnke},
  \citenamefont {Kunitski}, \citenamefont {He}, \citenamefont {Lein},
  \citenamefont {Eckart},\ and\ \citenamefont {D\"orner}}]{BurgdorferPRL2022}%
  \BibitemOpen
  \bibfield  {author} {\bibinfo {author} {\bibfnamefont {K.}~\bibnamefont
  {Lin}}, \bibinfo {author} {\bibfnamefont {S.}~\bibnamefont {Brennecke}},
  \bibinfo {author} {\bibfnamefont {H.}~\bibnamefont {Ni}}, \bibinfo {author}
  {\bibfnamefont {X.}~\bibnamefont {Chen}}, \bibinfo {author} {\bibfnamefont
  {A.}~\bibnamefont {Hartung}}, \bibinfo {author} {\bibfnamefont
  {D.}~\bibnamefont {Trabert}}, \bibinfo {author} {\bibfnamefont
  {K.}~\bibnamefont {Fehre}}, \bibinfo {author} {\bibfnamefont
  {J.}~\bibnamefont {Rist}}, \bibinfo {author} {\bibfnamefont {X.-M.}\
  \bibnamefont {Tong}}, \bibinfo {author} {\bibfnamefont {J.}~\bibnamefont
  {Burgd\"orfer}}, \bibinfo {author} {\bibfnamefont {L.~P.~H.}\ \bibnamefont
  {Schmidt}}, \bibinfo {author} {\bibfnamefont {M.~S.}\ \bibnamefont
  {Sch\"offler}}, \bibinfo {author} {\bibfnamefont {T.}~\bibnamefont {Jahnke}},
  \bibinfo {author} {\bibfnamefont {M.}~\bibnamefont {Kunitski}}, \bibinfo
  {author} {\bibfnamefont {F.}~\bibnamefont {He}}, \bibinfo {author}
  {\bibfnamefont {M.}~\bibnamefont {Lein}}, \bibinfo {author} {\bibfnamefont
  {S.}~\bibnamefont {Eckart}}, \ and\ \bibinfo {author} {\bibfnamefont
  {R.}~\bibnamefont {D\"orner}},\ }\href {\doibase
  10.1103/PhysRevLett.128.023201} {\bibfield  {journal} {\bibinfo  {journal}
  {Phys. Rev. Lett.}\ }\textbf {\bibinfo {volume} {128}},\ \bibinfo {pages}
  {023201} (\bibinfo {year} {2022}{\natexlab{a}})}\BibitemShut {NoStop}%
\bibitem [{\citenamefont {Lin}\ \emph {et~al.}(2022{\natexlab{b}})\citenamefont
  {Lin}, \citenamefont {Chen}, \citenamefont {Eckart}, \citenamefont {Jiang},
  \citenamefont {Hartung}, \citenamefont {Trabert}, \citenamefont {Fehre},
  \citenamefont {Rist}, \citenamefont {Schmidt}, \citenamefont {Sch\"offler},
  \citenamefont {Jahnke}, \citenamefont {Kunitski}, \citenamefont {He},\ and\
  \citenamefont {D\"orner}}]{PhysRevLett.128.113201}%
  \BibitemOpen
  \bibfield  {author} {\bibinfo {author} {\bibfnamefont {K.}~\bibnamefont
  {Lin}}, \bibinfo {author} {\bibfnamefont {X.}~\bibnamefont {Chen}}, \bibinfo
  {author} {\bibfnamefont {S.}~\bibnamefont {Eckart}}, \bibinfo {author}
  {\bibfnamefont {H.}~\bibnamefont {Jiang}}, \bibinfo {author} {\bibfnamefont
  {A.}~\bibnamefont {Hartung}}, \bibinfo {author} {\bibfnamefont
  {D.}~\bibnamefont {Trabert}}, \bibinfo {author} {\bibfnamefont
  {K.}~\bibnamefont {Fehre}}, \bibinfo {author} {\bibfnamefont
  {J.}~\bibnamefont {Rist}}, \bibinfo {author} {\bibfnamefont {L.~P.~H.}\
  \bibnamefont {Schmidt}}, \bibinfo {author} {\bibfnamefont {M.~S.}\
  \bibnamefont {Sch\"offler}}, \bibinfo {author} {\bibfnamefont
  {T.}~\bibnamefont {Jahnke}}, \bibinfo {author} {\bibfnamefont
  {M.}~\bibnamefont {Kunitski}}, \bibinfo {author} {\bibfnamefont
  {F.}~\bibnamefont {He}}, \ and\ \bibinfo {author} {\bibfnamefont
  {R.}~\bibnamefont {D\"orner}},\ }\href {\doibase
  10.1103/PhysRevLett.128.113201} {\bibfield  {journal} {\bibinfo  {journal}
  {Phys. Rev. Lett.}\ }\textbf {\bibinfo {volume} {128}},\ \bibinfo {pages}
  {113201} (\bibinfo {year} {2022}{\natexlab{b}})}\BibitemShut {NoStop}%
\bibitem [{\citenamefont {Peters}\ \emph {et~al.}(2022)\citenamefont {Peters},
  \citenamefont {Katsoulis},\ and\ \citenamefont
  {Emmanouilidou}}]{Agapi3electron}%
  \BibitemOpen
  \bibfield  {author} {\bibinfo {author} {\bibfnamefont {M.~B.}\ \bibnamefont
  {Peters}}, \bibinfo {author} {\bibfnamefont {G.~P.}\ \bibnamefont
  {Katsoulis}}, \ and\ \bibinfo {author} {\bibfnamefont {A.}~\bibnamefont
  {Emmanouilidou}},\ }\href {\doibase 10.1103/PhysRevA.105.043102} {\bibfield
  {journal} {\bibinfo  {journal} {Phys. Rev. A}\ }\textbf {\bibinfo {volume}
  {105}},\ \bibinfo {pages} {043102} (\bibinfo {year} {2022})}\BibitemShut
  {NoStop}%
\bibitem [{\citenamefont {Emmanouilidou}\ \emph {et~al.}(2023)\citenamefont
  {Emmanouilidou}, \citenamefont {Peters},\ and\ \citenamefont
  {Katsoulis}}]{AgapiNeonPRL}%
  \BibitemOpen
  \bibfield  {author} {\bibinfo {author} {\bibfnamefont {A.}~\bibnamefont
  {Emmanouilidou}}, \bibinfo {author} {\bibfnamefont {M.~B.}\ \bibnamefont
  {Peters}}, \ and\ \bibinfo {author} {\bibfnamefont {G.~P.}\ \bibnamefont
  {Katsoulis}},\ }\href {\doibase 10.48550/ARXIV.2302.03777} {\enquote
  {\bibinfo {title} {Singularity in electron-core potential as a gateway to
  accurate multi-electron ionization spectra in strongly driven atoms},}\ }
  (\bibinfo {year} {2023}),\ \Eprint {http://arxiv.org/abs/2302.03777}
  {arXiv:2302.03777 [physics.atom-ph]} \BibitemShut {NoStop}%
\bibitem [{\citenamefont {Ho}\ and\ \citenamefont
  {Eberly}(2006)}]{PhysRevLett.97.083001}%
  \BibitemOpen
  \bibfield  {author} {\bibinfo {author} {\bibfnamefont {P.~J.}\ \bibnamefont
  {Ho}}\ and\ \bibinfo {author} {\bibfnamefont {J.~H.}\ \bibnamefont
  {Eberly}},\ }\href {\doibase 10.1103/PhysRevLett.97.083001} {\bibfield
  {journal} {\bibinfo  {journal} {Phys. Rev. Lett.}\ }\textbf {\bibinfo
  {volume} {97}},\ \bibinfo {pages} {083001} (\bibinfo {year}
  {2006})}\BibitemShut {NoStop}%
\bibitem [{\citenamefont {Zhou}\ \emph {et~al.}(2010)\citenamefont {Zhou},
  \citenamefont {Liao},\ and\ \citenamefont {Lu}}]{Zhou:10}%
  \BibitemOpen
  \bibfield  {author} {\bibinfo {author} {\bibfnamefont {Y.}~\bibnamefont
  {Zhou}}, \bibinfo {author} {\bibfnamefont {Q.}~\bibnamefont {Liao}}, \ and\
  \bibinfo {author} {\bibfnamefont {P.}~\bibnamefont {Lu}},\ }\href {\doibase
  10.1364/OE.18.016025} {\bibfield  {journal} {\bibinfo  {journal} {Opt.
  Express}\ }\textbf {\bibinfo {volume} {18}},\ \bibinfo {pages} {16025}
  (\bibinfo {year} {2010})}\BibitemShut {NoStop}%
\bibitem [{\citenamefont {Tang}\ \emph {et~al.}(2013)\citenamefont {Tang},
  \citenamefont {Huang}, \citenamefont {Zhou},\ and\ \citenamefont
  {Lu}}]{Tang:13}%
  \BibitemOpen
  \bibfield  {author} {\bibinfo {author} {\bibfnamefont {Q.}~\bibnamefont
  {Tang}}, \bibinfo {author} {\bibfnamefont {C.}~\bibnamefont {Huang}},
  \bibinfo {author} {\bibfnamefont {Y.}~\bibnamefont {Zhou}}, \ and\ \bibinfo
  {author} {\bibfnamefont {P.}~\bibnamefont {Lu}},\ }\href {\doibase
  10.1364/OE.21.021433} {\bibfield  {journal} {\bibinfo  {journal} {Opt.
  Express}\ }\textbf {\bibinfo {volume} {21}},\ \bibinfo {pages} {21433}
  (\bibinfo {year} {2013})}\BibitemShut {NoStop}%
\bibitem [{\citenamefont {Kirschbaum}\ and\ \citenamefont
  {Wilets}(1980)}]{PhysRevA.21.834}%
  \BibitemOpen
  \bibfield  {author} {\bibinfo {author} {\bibfnamefont {C.~L.}\ \bibnamefont
  {Kirschbaum}}\ and\ \bibinfo {author} {\bibfnamefont {L.}~\bibnamefont
  {Wilets}},\ }\href {\doibase 10.1103/PhysRevA.21.834} {\bibfield  {journal}
  {\bibinfo  {journal} {Phys. Rev. A}\ }\textbf {\bibinfo {volume} {21}},\
  \bibinfo {pages} {834} (\bibinfo {year} {1980})}\BibitemShut {NoStop}%
\bibitem [{\citenamefont {Jiang}\ and\ \citenamefont
  {He}(2021)}]{PhysRevA.104.023113}%
  \BibitemOpen
  \bibfield  {author} {\bibinfo {author} {\bibfnamefont {H.}~\bibnamefont
  {Jiang}}\ and\ \bibinfo {author} {\bibfnamefont {F.}~\bibnamefont {He}},\
  }\href {\doibase 10.1103/PhysRevA.104.023113} {\bibfield  {journal} {\bibinfo
   {journal} {Phys. Rev. A}\ }\textbf {\bibinfo {volume} {104}},\ \bibinfo
  {pages} {023113} (\bibinfo {year} {2021})}\BibitemShut {NoStop}%
\bibitem [{\citenamefont {Jiang}\ \emph {et~al.}(2022)\citenamefont {Jiang},
  \citenamefont {Efimov}, \citenamefont {He},\ and\ \citenamefont
  {Prauzner-Bechcicki}}]{PhysRevA.105.053119}%
  \BibitemOpen
  \bibfield  {author} {\bibinfo {author} {\bibfnamefont {H.}~\bibnamefont
  {Jiang}}, \bibinfo {author} {\bibfnamefont {D.}~\bibnamefont {Efimov}},
  \bibinfo {author} {\bibfnamefont {F.}~\bibnamefont {He}}, \ and\ \bibinfo
  {author} {\bibfnamefont {J.~S.}\ \bibnamefont {Prauzner-Bechcicki}},\ }\href
  {\doibase 10.1103/PhysRevA.105.053119} {\bibfield  {journal} {\bibinfo
  {journal} {Phys. Rev. A}\ }\textbf {\bibinfo {volume} {105}},\ \bibinfo
  {pages} {053119} (\bibinfo {year} {2022})}\BibitemShut {NoStop}%
\bibitem [{\citenamefont {Pandit}\ \emph {et~al.}(2017)\citenamefont {Pandit},
  \citenamefont {Sentoku}, \citenamefont {Becker}, \citenamefont {Barrington},
  \citenamefont {Thurston}, \citenamefont {Cheatham}, \citenamefont {Ramunno},\
  and\ \citenamefont {Ackad}}]{Pandit2017}%
  \BibitemOpen
  \bibfield  {author} {\bibinfo {author} {\bibfnamefont {R.~R.}\ \bibnamefont
  {Pandit}}, \bibinfo {author} {\bibfnamefont {Y.}~\bibnamefont {Sentoku}},
  \bibinfo {author} {\bibfnamefont {V.~R.}\ \bibnamefont {Becker}}, \bibinfo
  {author} {\bibfnamefont {K.}~\bibnamefont {Barrington}}, \bibinfo {author}
  {\bibfnamefont {J.}~\bibnamefont {Thurston}}, \bibinfo {author}
  {\bibfnamefont {J.}~\bibnamefont {Cheatham}}, \bibinfo {author}
  {\bibfnamefont {L.}~\bibnamefont {Ramunno}}, \ and\ \bibinfo {author}
  {\bibfnamefont {E.}~\bibnamefont {Ackad}},\ }\href {\doibase
  10.1063/1.4990555} {\bibfield  {journal} {\bibinfo  {journal} {Phys.
  Plasmas}\ }\textbf {\bibinfo {volume} {24}},\ \bibinfo {pages} {073303}
  (\bibinfo {year} {2017})}\BibitemShut {NoStop}%
\bibitem [{\citenamefont {Pandit}\ \emph {et~al.}(2018)\citenamefont {Pandit},
  \citenamefont {Becker}, \citenamefont {Barrington}, \citenamefont {Thurston},
  \citenamefont {Ramunno},\ and\ \citenamefont {Ackad}}]{Pandit2018}%
  \BibitemOpen
  \bibfield  {author} {\bibinfo {author} {\bibfnamefont {R.~R.}\ \bibnamefont
  {Pandit}}, \bibinfo {author} {\bibfnamefont {V.~R.}\ \bibnamefont {Becker}},
  \bibinfo {author} {\bibfnamefont {K.}~\bibnamefont {Barrington}}, \bibinfo
  {author} {\bibfnamefont {J.}~\bibnamefont {Thurston}}, \bibinfo {author}
  {\bibfnamefont {L.}~\bibnamefont {Ramunno}}, \ and\ \bibinfo {author}
  {\bibfnamefont {E.}~\bibnamefont {Ackad}},\ }\href {\doibase
  10.1063/1.5024380} {\bibfield  {journal} {\bibinfo  {journal} {Phys.
  Plasmas}\ }\textbf {\bibinfo {volume} {25}},\ \bibinfo {pages} {043302}
  (\bibinfo {year} {2018})}\BibitemShut {NoStop}%
\bibitem [{\citenamefont {Landau}\ and\ \citenamefont
  {Lifshitz}(1965)}]{Landau}%
  \BibitemOpen
  \bibfield  {author} {\bibinfo {author} {\bibfnamefont {L.~D.}\ \bibnamefont
  {Landau}}\ and\ \bibinfo {author} {\bibfnamefont {E.~M.}\ \bibnamefont
  {Lifshitz}},\ }\href@noop {} {\emph {\bibinfo {title} {Quantum {M}echanics:
  {N}on-{R}elativistic {T}heory}}},\ \bibinfo {edition} {2nd}\ ed.\ (\bibinfo
  {publisher} {Pergamon},\ \bibinfo {address} {Oxford},\ \bibinfo {year}
  {1965})\BibitemShut {NoStop}%
\bibitem [{\citenamefont {Delone}\ and\ \citenamefont
  {Krainov}(1991)}]{Delone:91}%
  \BibitemOpen
  \bibfield  {author} {\bibinfo {author} {\bibfnamefont {N.~B.}\ \bibnamefont
  {Delone}}\ and\ \bibinfo {author} {\bibfnamefont {V.~P.}\ \bibnamefont
  {Krainov}},\ }\href {\doibase 10.1364/JOSAB.8.001207} {\bibfield  {journal}
  {\bibinfo  {journal} {J. Opt. Soc. Am. B}\ }\textbf {\bibinfo {volume} {8}},\
  \bibinfo {pages} {1207} (\bibinfo {year} {1991})}\BibitemShut {NoStop}%
\bibitem [{\citenamefont {Rubinstein}\ and\ \citenamefont
  {Froese}(2016)}]{ROTA1986123}%
  \BibitemOpen
  \bibfield  {author} {\bibinfo {author} {\bibfnamefont {R.~Y.}\ \bibnamefont
  {Rubinstein}}\ and\ \bibinfo {author} {\bibfnamefont {D.~P.}\ \bibnamefont
  {Froese}},\ }\href {\doibase 10.1002/9781118631980} {\emph {\bibinfo {title}
  {Simulation and the {M}onte {C}arlo {M}ethod}}},\ \bibinfo {edition} {3rd}\
  ed.\ (\bibinfo  {publisher} {Wiley},\ \bibinfo {address} {New Jersey},\
  \bibinfo {year} {2016})\BibitemShut {NoStop}%
\bibitem [{\citenamefont {HuP}\ \emph {et~al.}(1997)\citenamefont {HuP},
  \citenamefont {Liu},\ and\ \citenamefont {Chen}}]{HUP1997533}%
  \BibitemOpen
  \bibfield  {author} {\bibinfo {author} {\bibfnamefont {B.}~\bibnamefont
  {HuP}}, \bibinfo {author} {\bibfnamefont {J.}~\bibnamefont {Liu}}, \ and\
  \bibinfo {author} {\bibfnamefont {S.-G.}\ \bibnamefont {Chen}},\ }\href
  {\doibase https://doi.org/10.1016/S0375-9601(97)00811-6} {\bibfield
  {journal} {\bibinfo  {journal} {Phys. Lett. A}\ }\textbf {\bibinfo {volume}
  {236}},\ \bibinfo {pages} {533} (\bibinfo {year} {1997})}\BibitemShut
  {NoStop}%
\bibitem [{\citenamefont {Delone}\ and\ \citenamefont
  {Krainov}(1998)}]{Delone_1998}%
  \BibitemOpen
  \bibfield  {author} {\bibinfo {author} {\bibfnamefont {N.~B.}\ \bibnamefont
  {Delone}}\ and\ \bibinfo {author} {\bibfnamefont {V.~P.}\ \bibnamefont
  {Krainov}},\ }\href {\doibase 10.1070/pu1998v041n05abeh000393} {\bibfield
  {journal} {\bibinfo  {journal} {Phys.-Uspekhi}\ }\textbf {\bibinfo {volume}
  {41}},\ \bibinfo {pages} {469} (\bibinfo {year} {1998})}\BibitemShut
  {NoStop}%
\bibitem [{\citenamefont {Fechner}\ \emph {et~al.}(2014)\citenamefont
  {Fechner}, \citenamefont {Camus}, \citenamefont {Ullrich}, \citenamefont
  {Pfeifer},\ and\ \citenamefont {Moshammer}}]{PhysRevLett.112.213001}%
  \BibitemOpen
  \bibfield  {author} {\bibinfo {author} {\bibfnamefont {L.}~\bibnamefont
  {Fechner}}, \bibinfo {author} {\bibfnamefont {N.}~\bibnamefont {Camus}},
  \bibinfo {author} {\bibfnamefont {J.}~\bibnamefont {Ullrich}}, \bibinfo
  {author} {\bibfnamefont {T.}~\bibnamefont {Pfeifer}}, \ and\ \bibinfo
  {author} {\bibfnamefont {R.}~\bibnamefont {Moshammer}},\ }\href {\doibase
  10.1103/PhysRevLett.112.213001} {\bibfield  {journal} {\bibinfo  {journal}
  {Phys. Rev. Lett.}\ }\textbf {\bibinfo {volume} {112}},\ \bibinfo {pages}
  {213001} (\bibinfo {year} {2014})}\BibitemShut {NoStop}%
\bibitem [{\citenamefont {Montemayor}\ and\ \citenamefont
  {Schiwietz}(1989)}]{PhysRevA.40.6223}%
  \BibitemOpen
  \bibfield  {author} {\bibinfo {author} {\bibfnamefont {V.~J.}\ \bibnamefont
  {Montemayor}}\ and\ \bibinfo {author} {\bibfnamefont {G.}~\bibnamefont
  {Schiwietz}},\ }\href {\doibase 10.1103/PhysRevA.40.6223} {\bibfield
  {journal} {\bibinfo  {journal} {Phys. Rev. A}\ }\textbf {\bibinfo {volume}
  {40}},\ \bibinfo {pages} {6223} (\bibinfo {year} {1989})}\BibitemShut
  {NoStop}%
\bibitem [{\citenamefont {Heggie}(1974)}]{Heggie1974}%
  \BibitemOpen
  \bibfield  {author} {\bibinfo {author} {\bibfnamefont {D.~C.}\ \bibnamefont
  {Heggie}},\ }\href {\doibase 10.1007/BF01227621} {\bibfield  {journal}
  {\bibinfo  {journal} {Celest. Mech.}\ }\textbf {\bibinfo {volume} {10}},\
  \bibinfo {pages} {217} (\bibinfo {year} {1974})}\BibitemShut {NoStop}%
\bibitem [{\citenamefont {Pihajoki}(2015)}]{Pihajoki2015}%
  \BibitemOpen
  \bibfield  {author} {\bibinfo {author} {\bibfnamefont {P.}~\bibnamefont
  {Pihajoki}},\ }\href {\doibase 10.1007/s10569-014-9597-9} {\bibfield
  {journal} {\bibinfo  {journal} {Celest. Mech. Dyn. Astron.}\ }\textbf
  {\bibinfo {volume} {121}},\ \bibinfo {pages} {211} (\bibinfo {year}
  {2015})}\BibitemShut {NoStop}%
\bibitem [{\citenamefont {Liu}\ \emph {et~al.}(2016)\citenamefont {Liu},
  \citenamefont {Wu}, \citenamefont {Huang},\ and\ \citenamefont
  {Liu}}]{Liu2016}%
  \BibitemOpen
  \bibfield  {author} {\bibinfo {author} {\bibfnamefont {L.}~\bibnamefont
  {Liu}}, \bibinfo {author} {\bibfnamefont {X.}~\bibnamefont {Wu}}, \bibinfo
  {author} {\bibfnamefont {G.}~\bibnamefont {Huang}}, \ and\ \bibinfo {author}
  {\bibfnamefont {F.}~\bibnamefont {Liu}},\ }\href {\doibase
  10.1093/mnras/stw807} {\bibfield  {journal} {\bibinfo  {journal} {Mon. Not.
  R. Astron. Soc.}\ }\textbf {\bibinfo {volume} {459}},\ \bibinfo {pages}
  {1968} (\bibinfo {year} {2016})}\BibitemShut {NoStop}%
\bibitem [{\citenamefont {Katsoulis}\ \emph {et~al.}(2021)\citenamefont
  {Katsoulis}, \citenamefont {Peters}, \citenamefont {Staudte}, \citenamefont
  {Bhardwaj},\ and\ \citenamefont {Emmanouilidou}}]{PhysRevA.103.033115}%
  \BibitemOpen
  \bibfield  {author} {\bibinfo {author} {\bibfnamefont {G.~P.}\ \bibnamefont
  {Katsoulis}}, \bibinfo {author} {\bibfnamefont {M.~B.}\ \bibnamefont
  {Peters}}, \bibinfo {author} {\bibfnamefont {A.}~\bibnamefont {Staudte}},
  \bibinfo {author} {\bibfnamefont {R.}~\bibnamefont {Bhardwaj}}, \ and\
  \bibinfo {author} {\bibfnamefont {A.}~\bibnamefont {Emmanouilidou}},\ }\href
  {\doibase 10.1103/PhysRevA.103.033115} {\bibfield  {journal} {\bibinfo
  {journal} {Phys. Rev. A}\ }\textbf {\bibinfo {volume} {103}},\ \bibinfo
  {pages} {033115} (\bibinfo {year} {2021})}\BibitemShut {NoStop}%
\bibitem [{\citenamefont {Press}\ \emph {et~al.}(2007)\citenamefont {Press},
  \citenamefont {Teukolsky}, \citenamefont {Vetterling},\ and\ \citenamefont
  {Flannery}}]{press2007numerical}%
  \BibitemOpen
  \bibfield  {author} {\bibinfo {author} {\bibfnamefont {W.~H.}\ \bibnamefont
  {Press}}, \bibinfo {author} {\bibfnamefont {S.~A.}\ \bibnamefont
  {Teukolsky}}, \bibinfo {author} {\bibfnamefont {W.~T.}\ \bibnamefont
  {Vetterling}}, \ and\ \bibinfo {author} {\bibfnamefont {B.~P.}\ \bibnamefont
  {Flannery}},\ }\href
  {https://www.cambridge.org/gb/academic/subjects/mathematics/numerical-recipes/numerical-recipes-source-code-cd-rom-art-scientific-computing-3rd-edition-1?format=WW&isbn=9780521884075}
  {\emph {\bibinfo {title} {Numerical recipes: {T}he {A}rt of {S}cientific
  {C}omputing}}},\ \bibinfo {edition} {3rd}\ ed.\ (\bibinfo  {publisher}
  {Cambridge University Press},\ \bibinfo {address} {Cambridge},\ \bibinfo
  {year} {2007})\BibitemShut {NoStop}%
\bibitem [{\citenamefont {Bulirsch}\ and\ \citenamefont
  {Stoer}(1966)}]{bulirsch1966numerical}%
  \BibitemOpen
  \bibfield  {author} {\bibinfo {author} {\bibfnamefont {R.}~\bibnamefont
  {Bulirsch}}\ and\ \bibinfo {author} {\bibfnamefont {J.}~\bibnamefont
  {Stoer}},\ }\href {\doibase 10.1007/BF02165234} {\bibfield  {journal}
  {\bibinfo  {journal} {Numer. Math.}\ }\textbf {\bibinfo {volume} {8}},\
  \bibinfo {pages} {1} (\bibinfo {year} {1966})}\BibitemShut {NoStop}%
\bibitem [{\citenamefont {Abramowitz}\ and\ \citenamefont
  {Stegun}(1965)}]{abramowitz1965handbook}%
  \BibitemOpen
  \bibfield  {author} {\bibinfo {author} {\bibfnamefont {M.}~\bibnamefont
  {Abramowitz}}\ and\ \bibinfo {author} {\bibfnamefont {I.}~\bibnamefont
  {Stegun}},\ }\href {https://store.doverpublications.com/0486612724.html}
  {\emph {\bibinfo {title} {{H}andbook of {M}athematical {F}unctions: {W}ith
  {F}ormulas, {G}raphs, and {M}athematical {T}ables}}},\ Applied mathematics
  series\ (\bibinfo  {publisher} {Dover Publications},\ \bibinfo {year}
  {1965})\BibitemShut {NoStop}%
\end{thebibliography}%

\end{document}